\documentclass[10pt,prb,twocolumn,showpacs]{revtex4}
\usepackage{graphicx,amsmath,amssymb}
\begin{document}

\renewcommand{\Re}{\mathop{\mathrm{Re}}}
\renewcommand{\Im}{\mathop{\mathrm{Im}}}
\renewcommand{\b}[1]{\mathbf{#1}}
\renewcommand{\u}{\uparrow}
\renewcommand{\d}{\downarrow}
\newcommand{\bsigma}{\boldsymbol{\sigma}}
\newcommand{\blambda}{\boldsymbol{\lambda}}
\newcommand{\Tr}{\mathop{\mathrm{tr}}}
\newcommand{\sgn}{\mathop{\mathrm{sgn}}}
\newcommand{\sech}{\mathop{\mathrm{sech}}}
\newcommand{\diag}{\mathop{\mathrm{diag}}}
\newcommand{\half}{{\textstyle\frac{1}{2}}}
\newcommand{\sh}{{\textstyle{\frac{1}{2}}}}
\newcommand{\ish}{{\textstyle{\frac{i}{2}}}}
\newcommand{\thf}{{\textstyle{\frac{3}{2}}}}
\newcommand{\be}{\begin{equation}}
\newcommand{\ee}{\end{equation}}

\title{Models of three-dimensional fractional topological insulators}

\author{Joseph Maciejko$^1$, Xiao-Liang Qi$^2$, Andreas Karch$^3$ and Shou-Cheng Zhang$^2$}

\affiliation{$^1$Princeton Center for Theoretical Science \& Department of Physics,
Princeton University, Princeton, New Jersey 08544, USA\\
$^2$Department of Physics, Stanford
University, Stanford, California 94305, USA\\
$^3$Department of Physics, University of Washington, Seattle, Washington
98195-1560, USA}

\date\today

\begin{abstract}
Time-reversal invariant three-dimensional topological insulators can be defined fundamentally by a topological
field theory with a quantized axion angle $\theta$ of $0$ or $\pi$. It was recently shown that fractional quantized values of $\theta$ are consistent with time-reversal invariance if deconfined, gapped, fractionally charged bulk excitations appear in the low-energy spectrum due to strong correlation effects, leading to the concept of a fractional topological insulator. These fractionally charged excitations are coupled to emergent gauge fields which ensure that the microscopic degrees of freedom, the original electrons, are gauge-invariant objects. A first step towards the construction of microscopic models of fractional topological insulators is to understand the nature of these emergent gauge theories and their corresponding phases. In this work, we show that low-energy effective gauge theories of both Abelian or non-Abelian type are consistent with a fractional quantized axion angle if they admit a Coulomb phase or a Higgs phase with gauge group broken down to a discrete subgroup. The Coulomb phases support gapless but electrically neutral bulk excitations while the Higgs phases are fully gapped. The Higgs and non-Abelian Coulomb phases exhibit multiple ground states on boundaryless spatial $3$-manifolds with nontrivial first homology, while the Abelian Coulomb phase has a unique ground state. The ground state degeneracy receives an additional contribution on manifolds with boundary due to the induced boundary Chern-Simons term.
\end{abstract}


\maketitle

\section{Introduction}

Topological insulators\cite{qi2010a,hasan2010,qi2010} are new states of quantum matter that cannot be adiabatically connected to conventional insulators. They are fully gapped in the bulk but support gapless boundary modes which are protected by discrete symmetries.\cite{Qi2008,fu2007,moore2007,roy2009,schnyder2008,schnyder2009,kitaev2009,ryu2010} Topological insulators were first discovered in HgTe quantum wells.\cite{bernevig2006d,koenig2007} More recently, three-dimensional (3D) time-reversal invariant topological insulators have attracted a great deal of attention, for the most part due to the theoretical prediction\cite{fu2007b,zhang2009} and subsequent experimental detection\cite{Hsieh2008,Xia2009,Chen2009} of their protected helical surface states. The existence and stability of these surface states is protected by a bulk $\mathbb{Z}_2$ topological invariant which corresponds physically to a topological magnetoelectric effect\cite{Qi2008} (TME). The experimental observation of the TME, for instance by way of magnetooptical measurements,\cite{Qi2008,tse2010,maciejko2010} is a key goal in the field which is being actively pursued.\cite{laforge2010,butch2010,sushkov2010,jenkins2010,aguilar2011,hsieh2011} From a theoretical standpoint,\cite{Qi2008} the TME is described at energies much smaller than the energy gap by the addition of a term proportional to $\b{E}\cdot\b{B}$ (with $\b{E}$ and $\b{B}$ the electric and magnetic fields, respectively) to the usual Lagrangian for Maxwell electrodynamics, i.e., axion electrodynamics.\cite{Wilczek1987} This term is in fact the Abelian version of the topological $\theta$-term in quantum chromodynamics,\cite{jackiw1976,callan1976} and its coefficient $\theta$ is periodic (under certain conditions\cite{vazifeh2010}) with period $2\pi$. Since $\b{E}\cdot\b{B}$ is odd under time-reversal symmetry ($T$), the only values of $\theta$ allowed by $T$ are $0$ or $\pi$ mod $2\pi$, with $\theta=0$ for the $\mathbb{Z}_2$ trivial insulator and $\theta=\pi$ for the $\mathbb{Z}_2$ nontrivial insulator.\cite{Qi2008} Topological insulators can be described microscopically by noninteracting, spin-orbit coupled electrons hopping on a lattice, and the axion angle $\theta$ can be computed from a knowledge of the single-particle wave functions in momentum space.\cite{Qi2008,Essin2009}

In the presence of electron interactions, the concept of a topological insulator can no longer be defined in terms of band theory, but the fundamental definition in terms of the topological field theory and the axion response function remains generally valid.\cite{Qi2008} Weak interactions can either turn the noninteracting helical surface state into a weakly interacting helical Fermi liquid with spin-charge coupled collective modes,\cite{raghu2010} or drive a transition to a superconducting state.\cite{santos2010,cortijo2010} Strong enough interactions can lead to spontaneous $T$-breaking on the surface and ferromagnetic\cite{Qi2008,liu2009prl,xu2010,xu2010b,kim2010,ghaemi2010} or helical spin density wave order.\cite{jiang2010} In contrast, the bulk of a topological insulator is fully gapped and thus expected to be perturbatively stable to interactions. On the other hand, exotic states known as topological Mott insulators\cite{raghu2008,zhang2009b,kargarian2010,kurita2010} have been theoretically proposed, whereas a topologically nontrivial bandstructure is dynamically generated as a consequence of strong electron-electron interactions. Although these are strongly interacting states, their mean-field description is still that of a topological band insulator, and the axion angle $\theta$ remains quantized\cite{wang2010,gurarie2011} to $0$ or $\pi$ mod $2\pi$. Another type of topological Mott insulator has been theoretically proposed\cite{pesin2010,witczak-krempa2010} in which spin-charge separation\cite{SpinChargeMott} leads to a bulk insulator with a helical liquid of gapless spinons, but an electromagnetic $\theta$-term is not generated because the spinons are electrically neutral.

In many regards, the 3D $T$-invariant topological insulator can be viewed as a generalization of the 2D integer quantum Hall effect (IQHE) to 3D. The topological $\mathbb{Z}_2$ quantization of the bulk axion angle $\theta$ in 3D is the direct analog of the topological $\mathbb{Z}$ quantization of the bulk Hall conductance in 2D. By analogy with the relation between the IQHE and the fractional quantum Hall effect (FQHE), one is naturally led to the question whether there can exist a ``fractional 3D topological insulator'' which preserves $T$ but is characterized by a \emph{fractional} quantized axion angle, i.e., where $\theta$ is a non-integer, rational multiple of~$\pi$. A $T$-invariant fractional topological insulator was first proposed in two dimensions,\cite{bernevig2006a} and more recent works have shown the robustness of such a state.\cite{levin2009b,liu2009,Karch:2010mn} In the special case of conserved $S_z$, the QSH insulator is equivalent to two decoupled IQHE systems with equal and opposite effective magnetic fields. Both IQHE systems can be driven into FQHE states by adding electron-electron interactions and appropriately tuning the effective magnetic fields while keeping them equal and opposite, which yields a fractional QSH insulator without breaking $T$. On the other hand, this procedure is not directly applicable to three dimensions, where one cannot in general reduce a $T$-invariant topological insulator to two decoupled topological states which break $T$ in an equal and opposite way.\cite{3DFQHEnote}

Recently, a theory of $T$-invariant fractional topological insulators (fTI) in 3D was introduced\cite{Maciejko:2010tx,Swingle:2010rf} based on a parton construction.\cite{Jain1989,WenPartons} Postulating that an electron, under the influence of strong interactions in the underlying lattice Hamiltonian, fractionalizes into $N_c$ ``colors'' of partons gives a realization of a fractional topological insulator, as long as each color of partons forms a topological band insulator. To ensure that outside the fTI the partons recombine into electrons, one needs to introduce additional gauge degrees of freedom. Outside the topological insulator these emergent gauge fields are confining. As a consequence, the partons, which are charged under the emergent gauge fields, can never be observed in isolation but are instead confined into electrons, just like quarks are confined into mesons and baryons in quantum chromodynamics. The theory of the partons together with the emergent gauge fields should be thought of as a low-energy effective description of the system. The goal of this paper is to show that consistent low-energy theories with the characteristic properties of a fTI exist and to analyze their properties.\cite{noteLevinBurnellPaper}

There are many choices and questions associated with the additional emergent gauge fields. Are they Abelian or non-Abelian in nature? By construction they should confine outside the fTI, but what phase do they realize inside the fTI? As pointed out in Ref.~\onlinecite{Swingle:2010rf}, the gauge fields should be in a deconfined phase inside in the fTI, to ensure that the partons are propagating degrees of freedom. The search for deconfined phases in effective gauge theories of condensed matter systems and their identification as fractionalized phases of matter with unconventional types of order is an important question in modern theoretical condensed matter physics.\cite{WenBook} Deconfined $\mathbb{Z}_2$ gauge fields appear in effective theories of gapped spin liquids,\cite{read1989a,read1991,wen1991,mudry1994,moessner2001,moessner2001b} high-temperature superconductors,\cite{senthil2000,senthil2001} and non-Fermi liquids.\cite{ruegg2010,nandkishore2012} Deconfined $U(1)$ gauge fields appear in effective theories of gapped\cite{kalmeyer1987,wen1989,yao2007} and gapless\cite{moessner2003,hermele2004,lee2005} spin liquids, FQH liquids,\cite{zhang1989,zhang1992,read1989,blok1990} high-temperature superconductors,\cite{baskaran1988,affleck1988b,baskaran1989,lee2006} and unconventional quantum phase transitions.\cite{senthil2004,senthil2004b} Deconfined non-Abelian gauge fields such as $SU(N)$ gauge fields appear in parton constructions of FQH liquids.\cite{WenPartons,barkeshli2010} In the Abelian case, deconfined phases typically arise at small gauge coupling, with a spectrum consisting of weakly interacting gauge bosons which are gapped for discrete gauge groups such as the cyclic group $\mathbb{Z}_N$ and gapless for continuous gauge groups such as $U(1)$.\cite{WenBook} Deconfined phases of non-Abelian gauge fields are generally harder to come by, due to the renormalization group flow towards confinement at low energies (``infrared slavery'') which occurs even at weak coupling.\cite{GWP} A notable exception is the deconfining effect of the Chern-Simons term in $2+1$ dimensions even for non-Abelian gauge groups,\cite{pisarski1986,affleck1989,diamantini1993} which ensures the stability of non-Abelian parton constructions of FQH liquids.\cite{WenPartons,barkeshli2010} In Ref.~\onlinecite{Maciejko:2010tx} we mostly focused on the case of continuous gauge groups. In this case, it is easy to write down continuum gauge field theories with Abelian or non-Abelian groups and to demonstrate the theoretical consistency of fTIs in principle. Examples of both types were presented in our earlier work. One generic feature that all realizations of fTIs with continuous gauge groups share is the presence of additional gapless degrees of freedom. One typically considers topological insulator phases that are fully gapped in the bulk. In the case of continuous gauge groups we have to slightly generalize this understanding by demanding that all degrees of freedom charged under electromagnetism are gapped, while allowing for electrically neutral gapless degrees of freedom. In that case, the system is indeed an insulator as far as electrical transport is concerned. These additional neutral gapless degrees of freedom play a role similar to the low-energy phonons in a band insulator, and do not spoil the quantization of the electromagnetic response. We will refer to gapless deconfined phases as Coulomb phases.

Even if the fTI is described by an effective gauge theory with continuous gauge group at a given energy scale much less than the original electron bandwidth, it is possible that the system undergoes a transition to a Higgs phase at an even lower energy scale. We will present an explicit model of this kind in this work. The continuous gauge group is broken down to a discrete gauge group, and the latter is sufficient to ensure that in physical states the net electron number is integer. In this case, the system is truly gapped with no gapless degrees of freedom, but the partons are deconfined just as in the Coulomb phases.

As we will elaborate in more detail in this work, the basic topological feature of the fTI, a fractional $\theta$ angle, is completely robust in that it only depends on the total number of partons and not on any of the details of the emergent gauge sector --- whether it is Abelian or non-Abelian, and in a Coulomb or Higgs phase. This universal $\theta$ angle characterizes the TME in the material together with its physically measurable surface properties, such as a surface FQHE with half the conductivity of a typical Laughlin state per surface as well as the corresponding magnetooptical Kerr and Faraday effects.\cite{Qi2008,tse2010,maciejko2010,swingle2012}

There is, however, a second topological feature of the fTI that {\it does} depend on details of the gauge field sector: the ground state degeneracy. Some basic features of the ground state degeneracy have already been discussed in Ref.~\onlinecite{Maciejko:2010tx}, but here we will elaborate on this. It has recently been proven\cite{Swingle:2010rf} that for a fully gapped system, $\theta$ can only be fractional if the ground state on the $3$-torus $T^3$ is degenerate. This observation makes it clear that a confined phase is not an option for the fTI, as it would result in a unique ground state in a completely gapped system. In the case of a Coulomb realization of the fTI, we will find that the ground state in the Abelian models is unique, whereas in the non-Abelian models it is typically degenerate. The reason why this is not in contradiction with the theorem of Ref.~\onlinecite{Swingle:2010rf} can be traced to the fact that we have additional gapless degrees of freedom, therefore violating the assumption of Ref.~\onlinecite{Swingle:2010rf} that the system is gapped. We will demonstrate explicitly how this allows us to avoid the arguments of Ref.~\onlinecite{Swingle:2010rf}. In the Higgs models, the ground state is degenerate as the aforementioned theorem requires, and we determine the ground state degeneracy. The presence of either gapless degrees of freedom (for continuous gauge groups) or a finite topological degeneracy (for discrete gauge groups) is familiar from the study of topologically ordered spin systems, where the Lieb-Schultz-Mattis-Hastings theorem\cite{hastings2004} requires the existence of either type of low-energy modes.

The structure of the paper is as follows. In Sec.~\ref{SECPartonModels}, we introduce three general classes of parton-gauge boson effective theories: the Abelian $U(1)^{N_c-1}$ Coulomb models, the non-Abelian $SU(N_c)$ Coulomb models, and the Higgs $\mathbb{Z}_{N_c}$ models. We illustrate the general ideas in each case by a specific example of $N_c=3$ corresponding to three distinct classes of fTI but all with a fractional axion angle $\theta=\pi/3$. In Sec.~\ref{SECFQME}, we prove the topological quantization of the axion angle $\theta$ in fractional multiples of $\pi$. We derive the quantization of the axion angle in both Abelian and non-Abelian models, whether gapless or gapped. In Sec.~\ref{SECGSD}, we discuss the issue of ground state degeneracy on spatial $3$-manifolds of nontrivial topology. This is the direct analog of the topological degeneracy in FQH states on Riemann surfaces,\cite{topdegeneracy} and can be taken as an indication that fTI states exhibit topological order in the many-body sense.\cite{WenBook} In Sec.~\ref{SECSURFST} we briefly speculate on the nature of the gapless surface states, and summarize the paper in Sec.~\ref{SECCONCLUSION}. Appendix A and B clarify some technical issues concerning the ground state degeneracy on a $3$-manifold with boundary.%JM: I modified the last sentence to include the new appendix.

\section{Three classes of parton models}
\label{SECPartonModels}

In this section, we introduce the three basic classes of possible emergent gauge sectors for a fTI, and describe in more detail a specific $N_c=3$ model in each class.

\subsection{Coulomb models}
\label{Deconfined}

As described in the Introduction, one way to realize a fTI is to drive the emergent gauge fields into a Coulomb phase. This can be realized either in an Abelian or a non-Abelian setting with rather different properties. What is common to any Coulomb realization of a fTI is the appearance of extra gapless matter. This extra gapless matter is neutral from the point of view of the $U(1)_\mathrm{em}$ Maxwell gauge field. Therefore, the system is still an insulator, i.e., all degrees of freedom charged under $U(1)_\mathrm{em}$ are gapped. These additional gapless degrees of freedom can be considered as soft ``phonons'', since they do not enter the electrically charged sector.

\subsubsection{Abelian models}
\label{SecAbelianModels}

\begin{table}
\begin{tabular}{c||c|c|c}
field& $U(1)_\mathrm{em}$ & $U(1)_A$ & $U(1)_ B$ \\
\hline
$\psi_1$&$e/3$&$2g$&$-g$ \\
$\psi_2$&$e/3$&$-g$&$2g$\\
$\psi_3$&$e/3$&$-g$&$-g$\\
\end{tabular}
\caption{Gauge charge assignments of parton fields under electromagnetic $U(1)_\mathrm{em}$ and emergent $U(1)^2=U(1)_A\times U(1)_B$ gauge groups in the simplest Abelian model for $N_c=3$.}
\label{TableAbelian}
\end{table}
{\bf Model A:} In the simplest Abelian model of a fTI with $N_c=3$, the electron fractionalizes into three fermionic partons $\psi_i$, $i=1,2,3$ with an emergent $U(1)^2=U(1)_A\times U(1)_B$ gauge group in addition to the $U(1)_\mathrm{em}$ Maxwell gauge group. The parton Lagrangian is
\begin{align}\label{ModelALag}
\mathcal{L}=\psi_i^\dag\left(iD_0^{ij}-H_\theta(-i\b{D}^{ij})\right)\psi_j
+\mathcal{L}_\textrm{int}(\psi_i^\dag,\psi_i)+\mathcal{L}_\textrm{gf}(a_\mu),
\end{align}
where $H_\theta(\b{p})$ is the single-particle Hamiltonian for a noninteracting topological insulator with axion angle $\theta$, and
\begin{align}
D_\mu^{ij}=(D_0^{ij},-\b{D}^{ij})=\delta^{ij}\partial_\mu+iA_\mu t_\textrm{em}^{ij}+ia_\mu^a t_a^{ij}\nonumber
\end{align}
is the gauge-covariant derivative, with $A_\mu$ the electromagnetic gauge potential and $a_\mu=a_\mu^at_a$ the emergent gauge potential. $\mathcal{L}_\textrm{in}$ describes $T$-invariant residual interactions between the partons which do not destabilize their topological insulator ground state, and can thus be ignored. $\mathcal{L}_\textrm{gf}$ is the Lagrangian for the emergent gauge field $a_\mu$, and has to describe a Coulomb phase (e.g., the Maxwell Lagrangian). The $U(1)_\textrm{em}$ generator is $t_\mathrm{em}=\diag(e/3,e/3,e/3)$, and the $U(1)_A\times U(1)_B$ generators are $t_A=\diag(2g,-g,-g)$ and $t_B=\diag(-g,2g,-g)$, where $e$ is the electromagnetic gauge charge and $g$ is the emergent gauge charge. Equivalently, the charge assignments are given in Table~\ref{TableAbelian}. These gauge groups ensure that the only gauge-invariant operator that carries Maxwell charge is the product of the three parton operators, hence the gauge-invariant electron operator is $\psi_1 \psi_2 \psi_3$. Indeed, the generators of the weight lattice (see Sec.~\ref{quantNAmodels}) corresponding to the representation of Table~\ref{TableAbelian} are $\b{e}_1=(2g,-g)$, $\b{e}_2=(-g,2g)$, $\b{e}_3=(-g,-g)$, and the equation $\sum_i n_i\b{e}_i=0$ requiring an operator $\psi_1^{n_1}\psi_2^{n_2}\psi_3^{n_3}$ to be gauge-invariant has the one-parameter family of solutions $n_1=n_2=n_3$, i.e., operators of the form $(\psi_1\psi_2\psi_3)^n$ of which only $n=1$ corresponds to an operator with electromagnetic charge $e$. For a $T$-invariant system, all three partons have a real mass. As we will review in the next section, in the topologically nontrivial phase all three partons have a real but negative mass.

This model is the simplest of the $U(1)^{N_c}/U(1)_\mathrm{diag}$ Abelian models described in Ref.~\onlinecite{Maciejko:2010tx}. The generic model starts out with an emergent $U(1)^{N_c}$ gauge group and $N_c$ partons. Each parton carries charge $e/N_c$ under the Maxwell $U(1)_\mathrm{em}$ gauge group. The $i$th parton carries charge $g$ under the $i$th $U(1)$ factor of the emergent gauge group, and is neutral under the remaining $U(1)$ factors. One then takes the quotient of the emergent gauge group by its diagonal $U(1)_\mathrm{diag}$ subgroup whose generator is simply the sum of the generators of the individual $U(1)_i$ factors. For the particular case of $N_c=3$, the gauge group presented above represents the two remaining gauge group factors $U(1)_A = 2 U(1)_1 - U(1)_2 - U(1)_3$ and $U(1)_B = - U(1)_1 + 2 U(1)_2 - U(1)_3$. These are two linear combinations that are orthogonal to $U(1)_\mathrm{diag}$, in the sense that their generators satisfy $\Tr t_At_\mathrm{em}=\Tr t_Bt_\mathrm{em}=0$. For this choice of generators, one obtains the convenient feature that the emergent parton charges are integer multiples of $g$. These generators are however neither orthogonal to each other ($\Tr t_At_B\neq 0$) nor properly normalized ($\Tr t_{A,B}t_{A,B}\neq g^2$). Therefore, if one really starts out with a $U(1)^3$ gauge theory, the Maxwell gauge kinetic term $\Tr F_{\mu\nu}F^{\mu\nu}$ would contain a mixed $F^A_{\mu \nu} F_B^{\mu \nu}$ term. The generators can however be easily orthonormalized, and upon doing so we find precisely the generators $H_1$ and $H_2$ [Eq.~(\ref{CartanSU3})] of the maximal diagonal subgroup $U(1)^2$ of $SU(3)$ (see Sec.~\ref{quantNAmodels}). The same construction works for a general number of ``colors'' $N_c$, yielding once again parton charges which can be chosen to only take values $2g$ and $-g$ under the various Abelian factors, at the price of non-diagonal kinetic terms. As the gauge kinetic terms do not impact the topological properties this is of no importance. This allows one to study a simplified, topologically equivalent version of the model where one takes the above charge assignments with standard Maxwell terms (no mixing) and forgets about the fact that the $U(1)^{N_c-1}$ gauge group originated from $U(1)^{N_c}$ in the first place. If desired, a set of orthonormal generators is provided by the Cartan generators $H_1,\ldots,H_{N_c-1}$ of $SU(N_c)$, i.e., the generators of its maximal diagonal subgroup $U(1)^{N_c-1}$. In fact, as far as the electrically charged degrees of freedom are concerned, the $U(1)^{N_c}/U(1)_\mathrm{diag}$ model discussed above is equivalent to a non-Abelian $SU(N_c)$ model where $SU(N_c)$ is spontaneously broken to its maximal diagonal subgroup $U(1)^{N_c-1}$.

Upon integrating out the massive partons and setting the electromagnetic gauge potential to zero, in the fTI phase we are left with a pure $U(1)^{N_c-1}$ dynamical gauge theory with $\theta=\pi$ mod $2\pi$. For a continuum theory, there is a single phase, the deconfined Coulomb phase. This phase corresponds to a free field infrared fixed point at which the renormalized coupling $g$ vanishes. In other words, in the infrared we have free Dirac fermions (the partons) and photons (the emergent gauge fields). For a theory defined on the lattice, we obtain two phases, the deconfined Coulomb phase and the confined phase.\cite{wilson1974,polyakov1975,banks1977} The fTI phase corresponds to the Coulomb phase, which is equivalent to the continuum theory in the infrared except for a doubling of the number of fermion species. The topological properties are the same in both the continuum and lattice cases. However, the value ($0$ or $\pi$ mod $2\pi$) of the $\theta$ angle for the (free) partons is dependent upon the choice of regularization procedure in the continuum theory, while it is fixed by the lattice Hamiltonian in the lattice theory.\cite{Qi2008} In analogy with the results of Cardy and Rabinovici\cite{cardy1982} and Cardy\cite{cardy1982b} for $\mathbb{Z}_N$ gauge theory in $3+1$ dimensions, one may however wonder whether the emergent $\theta$-term with nonzero $\theta=\pi$ could affect the phase diagram of the emergent gauge sector, since the gauge theory is now dynamical. This is not so because a $U(1)$ gauge theory can be viewed as the $N\rightarrow\infty$ limit of a $\mathbb{Z}_N$ gauge theory, and in this limit the ``electric'' charges of $\mathbb{Z}_N$ gauge theory disappear from the spectrum\cite{ukawa1980} and the $\theta$-term has no effect on the bulk free energy.

\subsubsection{Non-Abelian models}

One may question whether a ``Coulomb'' phase is actually realizable for non-Abelian models, given the infrared renormalization group flow towards strong coupling in pure $SU(N_c)$ Yang-Mills theory. Although the latter theory is generally believed to be confining at zero temperature, adding gapless matter can result in a deconfined phase. For example, the flow to strong coupling is reversed if a large enough number of flavors of gapless fermions couple to the $SU(N_c)$ gauge field.\cite{GWP} Another example is given by $\mathcal{N}=4$ supersymmetric Yang-Mills (SYM) theory with any gauge group, in particular with gauge group $SU(N_c)$, which is conformally invariant and hence in a Coulomb phase.\cite{seiberg1988} In fact, this is the generic case even in non-supersymmetric Yang-Mills theories with enough gapless matter: for matter in the fundamental representation as well as for matter in two-index tensor representations it is believed that $SU(m)$, $SO(m)$ and $Sp(m)$ gauge theories all exhibit a ``conformal window", that is, the gauge theory flows in the infrared to a conformally invariant stable fixed point corresponding to a strongly coupled but deconfined phase as long as the number of flavors $N_f$ is within a certain finite range. A conjecture for the exact values of $N_f$ that bound the conformal window has been put forward for example in Ref.~\onlinecite{Dietrich:2006cm}. The upper end of the conformal window is theoretically well established: for $N_f$ larger than this maximal value the gauge theory loses asymptotic freedom, i.e., it flows to the Gaussian fixed point at low energies. Just below the upper end of the conformal window it can be established, using perturbation theory, that at least for large $N_c$ the theory indeed flows at low energies to a stable conformal fixed point known as the Banks-Zaks fixed point.\cite{Banks:1981nn} The lower bound of the conformal window is mainly conjectural based on partial resummations. There has been a lot of recent activity on numerical studies of the conformal window using lattice gauge theory.\cite{DeGrand:2010ba} While the precise lower bound of the conformal window is still up to debate, the existence of a conformal window has been firmly established, even for non-supersymmetric gauge theories. In gauge theories with $\mathcal{N}=1$ supersymmetry the full conformal window has been mapped out using the power of holomorphy.\cite{Seiberg:1994pq} For example, the $\mathcal{N}=1$ supersymmetric $SU(N_c)$ gauge theory flows in the infrared to a stable, deconfined, conformal fixed point for $\frac{3}{2} N_c < N_f < 3 N_c$.

As for the Abelian models, those Coulomb non-Abelian models have additional gapless degrees of freedom. For example, in the case of $\mathcal{N}=4$ SYM theory the extra matter consists of adjoint fermions and scalars, in addition to the gauge fields. One important difference compared to the Abelian case is that in the case of a non-Abelian gauge theory at a nontrivial fixed point the extra matter is not free, but remains interacting with a fixed renormalized coupling $g_*$, where $g_*$ is fixed by the requirement that the renormalization group $\beta$-function vanishes. In the special case of $\mathcal{N}=4$ SYM theory, $g_*$ is in fact a free parameter and the $\beta$-function vanishes identically for all values of the coupling.\cite{seiberg1988} Not only do the emergent gauge fields remain strongly coupled to each other, the partons remain strongly coupled to the emergent gauge fields. Below we will argue that the value of the fractional $\theta$ angle only depends on the number of partons the electron fractionalizes into, and is completely robust even against these strong interactions that remain in the low-energy effective theory of the partons. In fact, in case of $\mathcal{N}=4$ SYM theory this has been demonstrated explicitly in the extreme limit of very large coupling, employing a holographic realization of this particular model.\cite{HoyosBadajoz:2010ac}

{\bf Model B:} The canonical example in this class, which we will refer to as an example in various sections, is a model with $N_c=3$ partons of electric charge $e/3$ coupled in the fundamental representation to a $SU(3)$ gauge field with some additional massless adjoint matter fields which we denote collectively by $\Psi^a$,
\begin{eqnarray}\label{ModelBLag}
\mathcal{L}&=&\psi_i^\dag\left(iD_0^{ij}-H_\theta(-i\b{D}^{ij})\right)\psi_j+\mathcal{L}_\textrm{int}(\psi_i^\dag,\psi_i)\nonumber\\
&&+\mathcal{L}_\textrm{gf}(a_\mu)+\mathcal{L}_\textrm{matter}(\Psi^a,\partial_\mu\Psi^a-gf^{abc}a_\mu^b\Psi^c),
\end{eqnarray}
where $f^{abc}$ are the structure constants of $SU(3)$.\cite{Georgi:1982jb} The role of the matter fields $\Psi^a$, which couple only to $a_\mu$ and not to $A_\mu$, is to drive $a_\mu$ into a Coulomb phase. The detailed form of $\mathcal{L}_\textrm{matter}$ is largely irrelevant from a topological point of view, except for the question of the ground state degeneracy on $T^3$. $\mathcal{N}=4$ SYM theory can serve as the canonical example. It should be clarified that the existence of massless matter fields is consistent with the partons being gapped. The massive partons, which carry the fundamental representation of the $SU(3)$ group, are the only electrically charged particles and dominate electromagnetic transport. Being gapped, they are irrelevant for the low-energy dynamics of the emergent gauge field. The latter is driven by the massless $SU(3)$ charged but electrically neutral extra matter. %XLnote: I added the last sentence. Pls see if it's ok. %AK changed a little %JM: fixed typo

One consequence of the residual interactions at low energies is that sometimes $T$ is spontaneously broken in the topologically nontrivial phase, that is at $\theta=\pi$. Whether this happens or not depends on many details of the emergent gauge sector.\cite{Vicari:2008jw} For the purpose of constructing parton models of fTI, the important lesson to remember is that, even though spontaneous $T$-breaking does sometimes occur, Coulomb phases of non-Abelian gauge theories with unbroken $T$ do exist. In the large $N_c$ limit the situation is slightly better understood:\cite{Witten:1998uka} while pure $SU(N_c)$ Yang-Mills theory is believed to spontaneously break $T$ at $\theta=\pi$, additional gapless matter can prevent this. In particular, $\mathcal{N}=4$ $SU(N_c)$ SYM theory is $T$-invariant at $\theta=\pi$ in the large $N_c$ limit.

One other important difference between the Abelian and the non-Abelian models concerns spin. In model A, each of the three partons belongs to a different one-dimensional irreducible representation of the emergent gauge group $U(1)^2$. The requirement that for a fermionic state the many-body wave function has to be antisymmetric under interchange of both color and spin indices does not constrain the symmetrization properties of the spin quantum numbers, and hence the total spin, because gauge invariance does not require a separate, complete antisymmetrization of the color indices in this case. A spin-$1/2$ electron is always possible. In model B however, gauge invariance under the $SU(3)$ group requires that the many-body wave function be completely antisymmetric in color indices. Correspondingly the spin indices have to be completely symmetrized, and in model B the electron would have spin $3/2$. Because spin rotation invariance is already broken in a topological band insulator by spin-orbit coupling, this does not create a problem, but it is certainly an aspect of our non-Abelian models to keep in mind.

\subsubsection{General Coulomb model}
\label{generalmodel}

The general deconfined model will have some Abelian (free) factors and some non-Abelian (interacting) factors. A large class of models of this type was introduced in our earlier work.\cite{Maciejko:2010tx} The emergent gauge group in this general model is $\prod_{a=1}^{N_f} U(N_c^{a})/U(1)_\mathrm{diag}$. The total number of partons is
\be
N_c = \sum_{a=1}^{N_f} \, N_c^a.
\ee
For every flavor $a=1,\ldots,N_f$ we have $N_c^{a}$ partons transforming in the fundamental representation of the non-Abelian $SU(N_c^{a})$ factor and carrying $U(1)_a$ charge $q^a = g/N_c^a$ as well as electromagnetic charge $q^\mathrm{em}_a$, while being neutral under all the other gauge groups. The diagonal subgroup $U(1)_\mathrm{diag}$, whose generator is the sum of all the $U(1)^a$ generators, is modded out. The only gauge-invariant operator that carries nonvanishing electromagnetic charge is the product of all the partons. To be gauge-invariant under each of the individual non-Abelian factors, one needs to form baryonic operators out of the partons in the fundamental representation of that factor. These individual baryons however will carry $U(1)_a$ charge $g$. Only the product of all the individual baryons is gauge invariant, as this is the only way to get a gauge-invariant operator whose only emergent $U(1)$ charge is the charge under the diagonal subgroup $U(1)_\mathrm{em}$, which is removed from the emergent gauge group. The electromagnetic charge of this gauge-invariant operator is
\be
\label{chargeconstraint}
Q = \sum_{a=1}^{N_f} N_c^{a} q^\mathrm{em}_a,
\ee
hence the $q^\mathrm{em}_a$ have to be chosen in such a way that $Q=e$.

An alternative to $SU(N_c)$ gauge groups is to use models based on orthogonal or symplectic gauge groups. The gauge-invariant operators in such theories are the mesons $q_a q_b \delta^{ab}$ and $q_a q_b \omega^{ab}$, respectively, where $\delta^{ab}$ is the Kronecker delta and $\omega^{ab}$ the corresponding antisymmetric invariant tensor for the symplectic group. Both theories have baryons $q_{a_1} \cdots q_{a_{N_c}}
\epsilon^{a_1 \cdots a_{N_c}}$. However, in the $Sp(N_c)$ theory where $N_c$ is even, these baryons are not independent operators but are equivalent to a product of mesons, because the symplectic tensor $\omega^{ab}$ is antisymmetric. By analogy with the $[U(M)\times Sp(2k)]_1$ Chern-Simons theory of the $\mathbb{Z}_k$ parafermion FQH states obtained from a parton construction,\cite{barkeshli2010} these theories hold the promise of generating more exotic surface states. However, if we take the partons to have electric charge $q$, not only do we obtain baryons with electric charge $N_c q$ which would suggest $q=e/N_c$ as before, but the mesons are also charged as we can make mesons from two fundamental quarks. In the $SU(N_c)$ case, mesons are made from a fundamental quark and an antifundamental antiquark. Therefore, this time we are forced to identify the mesons with the electrons.\cite{barkeshli2010} Furthermore, we have to assign the partons charge $q/2$, and at the same time take the number of partons $N_c$ to be even so that the baryons carry an integer multiple of the electron charge, which is consistent with the fact that symplectic groups are only defined for even $N_c$. To ensure that this mesonic electron is a fermion, we need the electron to split into two different partons, a fermion and a boson, both in the fundamental representation of the gauge group. We would also get gauge-invariant scalar bosons with the same charge $e$ as the electron, but presumably these can be made gapped. In this case, the axion angle $\theta$ has to be an integer multiple of $N_c \pi/4$ from the $N_c$ fermions of charge $e/2$. This structure is somewhat reminiscent of the $\mathbb{Z}_2$ spin liquid model of a fTI put forward in Ref.~\onlinecite{Swingle:2010rf}.

\subsection{Higgs models}
\label{HiggsModels}

In order to get a completely gapped system that realizes a topological insulator, it is sufficient to add electrically neutral Higgs fields to the Coulomb models of Sec.~\ref{Deconfined}, and consider a specific pattern of spontaneous symmetry breaking $G\rightarrow H$ with $G$ the original gauge group, such that the unbroken gauge group $H$ in the Higgs phase is discrete. One can modify model B above to realize this possibility:

{\bf Model C:} A simple Higgs model can be obtained by augmenting the simplest non-Abelian three-parton model, model B above [Eq.~(\ref{ModelBLag})], by inclusion of two complex scalar Higgs fields $\phi_1^a$ and $\phi_2^a$ in the adjoint representation of $SU(3)$,
\begin{eqnarray*}
\mathcal{L}&=&\psi_i^\dag\left[iD_0^{ij}-H_\theta(-i\mathbf{D}^{ij})\right]\psi_j
+\mathcal{L}_\textrm{int}(\psi_i^\dag,\psi_i)\\
&&+\mathcal{L}_\textrm{gf}(a_\mu)+\mathcal{L}_\textrm{matter}(\Psi,\mathcal{D}_\mu\Psi)\\
&&+{\textstyle\frac{1}{2}}(\mathcal{D}_\mu\phi_1)^\dag
\mathcal{D}^\mu\phi_1
+{\textstyle\frac{1}{2}}(\mathcal{D}_\mu\phi_2)^\dag
\mathcal{D}^\mu\phi_2\nonumber\\
&&-V_1(\phi_1^\dag\phi_1)-V_2(\phi_2^\dag\phi_2),
\end{eqnarray*}
where $(\mathcal{D}_\mu\Psi)^a=\partial_\mu\Psi^a-gf^{abc}a_\mu^b\Psi^c$, and similarly for $(\mathcal{D}_\mu\phi_1)^a$ and $(\mathcal{D}_\mu\phi_2)^a$, is the gauge-covariant derivative with respect to the $SU(3)$ gauge field $a_\mu$ alone in the adjoint representation. The potentials $V_1$ and $V_2$ are $SU(3)$ invariant.
\begin{table}\label{TableHiggs}
\begin{tabular}{c||c|c}
field& $U(1)_\mathrm{em}$ & $\mathbb{Z}_3$  \\
\hline
$\psi_1$&$e/3$&1\\
$\psi_2$&$e/3$&1\\
$\psi_3$&$e/3$&1\\
\end{tabular}
\caption{Gauge charge assignments of parton fields under electromagnetic $U(1)_\mathrm{em}$ and unbroken emergent $\mathbb{Z}_3$ gauge groups in the simplest Higgs model for $N_c=3$. The third column corresponds to the triality of the representation.}
\end{table}
The center of $SU(3)$ is $\mathbb{Z}_3$ and the adjoint Higgs fields $\phi_1$ and $\phi_2$ are neutral under this center symmetry. On the one hand, if (say) only $V_1$ is such that $\langle\phi_1\rangle\neq 0$ while $\langle\phi_2\rangle=0$, we can always perform a global $SU(3)$ transformation to diagonalize $\langle\phi_1\rangle$. The continuous part of the gauge group is broken to its maximal diagonal subgroup $U(1)^2$ and one recovers the fermions and gauge fields of model A together with several charged scalars. On the other hand, if $V_1$ and $V_2$ are such that $\langle\phi_1\rangle$ and $\langle\phi_2\rangle$ acquire generic noncommuting expectation values, the continuous part of the gauge group is completely broken and generically only the discrete $\mathbb{Z}_3$ center symmetry is unbroken. Under this discrete gauge group the parton charges are given in Table~\ref{TableHiggs}. The third column corresponds to the nonzero triality $k=1$~mod~$3$ of the fundamental representation with character $e^{2\pi i/3}$. In general, for a $SU(N_c)\rightarrow\mathbb{Z}_{N_c}$ Higgs mechanism we want the partons to be in a representation of $SU(N_c)$ with nonzero $N_c$-ality $k=1,2,\ldots,N_c-1$~mod~$N_c$ with character $e^{2\pi ik/N_c}\neq 1$. This discrete subgroup of the original continuous emergent gauge group is completely sufficient to ensure that all gauge-invariant operators have charges that are an integer multiple of the electron charge. Indeed, the center $\mathbb{Z}_3$ of $SU(3)$ is generated by a single element of the Cartan subalgebra, $H_2\propto\diag(1,1,-2)$ in Eq.~(\ref{CartanSU3}). The weight lattice of the emergent gauge field is generated by the (unnormalized) weight vectors $\b{e}_1=1$, $\b{e}_2=1$, $\b{e}_3=-2$. An operator $\psi_1^{n_1}\psi_2^{n_2}\psi_3^{n_3}$ neutral with respect to the emergent gauge field must satisfy $\sum_in_i\b{e}_i=0$ which leads to $n_1+n_2=2n_3$, hence the total Maxwell charge is $e(n_1+n_2+n_3)/3=n_3e\in\mathbb{Z}e$. $\psi_1 \psi_2 \psi_3$ is still the simplest gauge-invariant operator, but as long as the fermions have internal spin states that can be antisymmetrized, $\psi_i^3$ or $\psi_i^2 \psi_j$ with $i,j=1,2,3$ and $i \neq j$ would give rise to additional gauge-invariant operators. Physically these operators can be considered as an electron combined with condensed Higgs bosons. %XLnote: a sentence is added here.
In the relativistic continuum theory the fermions carry spin $1/2$ and $\psi_i^3$ vanishes identically. As the fundamental degrees of freedom of the theory are the partons, the presence of these extra bound states carrying integer electron charge $e$ does not affect the physics. The important requirement is that Gauss' law, which enforces overall $\mathbb{Z}_3$ neutrality, ensures that the net charge of the whole sample is an integer multiple of the electron charge.

For Higgs fields that transform trivially under the center $\mathbb{Z}_{N_c}$ of $SU(N_c)$, as is the case here, it is known that the confined and Higgs phases are distinct.\cite{fradkin1979} In the Higgs phase, the system is completely gapped even in the charge neutral sector because the unbroken gauge group $\mathbb{Z}_3$ is discrete. However, in the limit of infinite Higgs stiffness $\kappa\rightarrow\infty$ the system behaves like a $\mathbb{Z}_{N_c}$ gauge theory which has an additional gapless Coulomb phase separating the (gapped) Higgs and confined phases for large enough $N_c$.\cite{fradkin1979,ukawa1980} It it possible that this phase persists for finite but large enough $\kappa$.\cite{ukawa1980} In this case, there would be two distinct fTI phases with the same value of $\theta$, one gapless and the other fully gapped. We also note that the presence of a $\theta=\pi$ term in the emergent $\mathbb{Z}_{N_c}$ gauge theory gives rise to the presence of oblique confined phases with dyon condensation in addition to the usual confined phase,\cite{cardy1982,cardy1982b} but does not remove the Higgs and Coulomb phases corresponding to the fTI. Finally, one can also construct Abelian Higgs models in addition to the non-Abelian Higgs model discussed here, with $U(1)^{N_c-1}$ broken to a discrete subgroup. In the simplest case $N_c=3$, we can add two charge-3 complex scalars $\phi_1$, $\phi_2$ to the Lagrangian (\ref{ModelALag}) of Model A,
\begin{eqnarray*}
\mathcal{L}&=&\psi_i^\dag\left(iD_0^{ij}-H_\theta(-i\b{D}^{ij})\right)\psi_j+\mathcal{L}_\textrm{int}(\psi_i^\dag,\psi_i)+\mathcal{L}_\textrm{gf}(a_\mu)\\
&&+\half|(\partial_\mu+3iga_\mu^1)\phi_1|^2+\half|(\partial_\mu+3iga_\mu^2)\phi_2|^2\\
&&-V_1(|\phi_1|^2)-V_2(|\phi_2|^2),
\end{eqnarray*}
which is invariant under the $U(1)^2$ gauge transformations $\psi\rightarrow e^{i\Lambda^a t_a}\psi$ (sum over $a$), $\phi_a\rightarrow e^{3ig\Lambda^a}\phi_a$ (no summation), $a_\mu^a\rightarrow a_\mu^a-\partial_\mu\Lambda^a$. If $V_1$ and $V_2$ are such that $\langle\phi_1\rangle\neq 0$ and $\langle\phi_2\rangle\neq 0$, the only gauge transformations which leave the vacuum invariant are $\Lambda^1=\frac{2\pi k_1}{3g}$, $\Lambda^2=\frac{2\pi k_2}{3g}$, $k_1,k_2=0,1,2$ mod 3. From the point of view of the partons, these correspond to $e^{i\Lambda^at_a}=e^{-2\pi ik/3}$, $k=k_1+k_2=0,1,2$ mod $3$ which form the discrete group $\mathbb{Z}_3$. It has been shown recently\cite{banks2011} that the topological limit of $\mathbb{Z}_{N_c}$ gauge theory in $3+1$ dimensions is described by a level-$N_c$ $BF$ theory. In that sense, the $\mathbb{Z}_{N_c}$ fTI discussed here is similar to the fTI discussed in Sec. 6 of Ref.~\onlinecite{cho2011}.

We conclude this section with a brief discussion of the extent to which the low-energy effective theories we have discussed so far have a chance of being realized in a microscopic model of electrons. The Abelian $U(1)^{N_c-1}$ theory contains only partons and gauge fields which arise naturally from the parton decomposition of the electron. Therefore, there is no principle which forbids the realization of such a field theory in an electron model. As discussed, the non-Abelian $SU(N_c)$ models typically require extra gapless electrically neutral fermions to achieve a Coulomb phase. It is not at present clear to us how such low-energy effective theories could arise from an electron model, since charge $e$ bosons could emerge as bound states of an electron and a neutral fermion. We speculate that the non-Abelian models might be relevant for cold atom systems where charge $e$ fermions and charge $e$ bosons can occur. On the other hand, the Higgs phases of the non-Abelian models are possible in a pure electron model, with the $SU(N_c)$ gauge group spontaneously broken to a subgroup such as $U(1)^{N_c-1}$ or $\mathbb{Z}_{N_c}$ due to the condensation of parton bilinears (i.e., ``composite Higgs'') of the form $\phi^a=\psi_i^\dag t_a^{ij}\psi_j$ with $t_a^{ij}$ the $SU(N_c)$ generators, i.e., ``color'' superconductivity in the parton sector. Note that a ground state with $\langle\phi^a\rangle\neq 0$ would carry a ``color'' supercurrent but no electromagnetic supercurrent, hence the electromagnetic response would truly be that of a fractional axion insulator rather than that of a superconductor.

\section{Fractional topological magnetoelectric effect}
\label{SECFQME}

\subsection{Chiral anomaly}

The calculation of the effective axion angle $\theta$ via the chiral or Adler-Bell-Jackiw (ABJ) anomaly\cite{Adler:1969gk,Bell:1969ts} was mentioned in our original work\cite{Maciejko:2010tx} following earlier discussions,\cite{Qi2008} and has been spelled out in more detail elsewhere.\cite{hosur2010,ryu2010,HoyosBadajoz:2010ac,Karch:2010mn,Ryu2010b,Mulligan:2010hi} This argument can be used to obtain topological insulators in any even spacetime dimension,\cite{Qi2008,ryu2010,Karch:2010mn,Mulligan:2010hi} but for here let us specialize to the case of the 3D $T$-invariant topological insulator. The goal is to calculate the contribution to the effective $\theta$ angle of a $U(1)$ gauge group that arises from integrating out a fermion of charge $q$. As (at least in model A) our partons are charged under more than one Abelian group, we want to calculate all terms of the form
\be
\label{stheta}
 S_\theta = \frac{ i\sum_{a,b} \theta_{ab} e^2}{32\pi^2}\int_\mathcal{M} d^4x\,\epsilon_{\mu\nu\lambda\rho}F^a_{\mu\nu}F^b_{\lambda\rho},
\ee
where $\mathcal{M}$ is the (here, Euclidean) spacetime manifold and the label $a$ runs over all the Abelian groups in the problem, i.e., the emergent gauge groups as well as the Maxwell gauge group. For example, in model A we have $a,b\in\{\mathrm{em},A,B\}$. The non-Abelian case works similarly as will be discussed below. Let us denote by $q_i^a$ the charge of the $i$th parton under the $a$th gauge group in units of the corresponding gauge coupling ($e$ for the electromagnetic sector, $g$ for the emergent sector). In the Dirac kinetic term for the parton, one can write a complex mass term with mass $M$ as the complex bilinear operator $\overline{\psi} ( \Re M + i\gamma_5\Im M) \psi$. The $T$ operator takes $M$ into its complex conjugate $M^*$, so the system is only $T$-invariant if $M$ is real. In other even spacetime dimensions it is a different discrete symmetry which takes the role of enforcing a real mass term.\cite{ryu2010} Once $M$ is real we see that there is a $\mathbb{Z}_2$ choice of mass terms: $M$ can be real and positive or real and negative. The two can not be smoothly deformed into each other in a $T$-invariant fashion without crossing $M=0$, that is, without closing the gap.

As all that matters in terms of physics of interfaces is the difference in $\theta$, we can choose the $\theta$-term to be zero in the case that $M$ is real and positive, which one identifies as the topologically trivial case. At the classical level, we can always perform a chiral rotation $\psi \rightarrow e^{i \alpha \gamma_5} \psi$ to absorb the phase of $M$. This chiral rotation is a symmetry of the massless theory. One can think of the mass as a spurion, i.e., as arising from the expectation value of a non-dynamical background field, to restore the symmetry in the massive case by letting $M$ transform as $M \rightarrow e^{2 i \alpha} M$. If the phase of $M$ is originally $\theta_0$, with $\theta_0=\pi$ for the topologically nontrivial $T$-invariant insulator, $M$ can be made real and positive by a chiral rotation with angle $\alpha = - \theta_0/2$. However, this chiral rotation is not a symmetry of the quantum effective action, as the path integral measure is not invariant. Performing such a rotation generates a $\theta$-term of the form given in Eq.~($\ref{stheta}$), with a coefficient that is determined by a triangle diagram with one axial current and two $U(1)$ currents,
\be
\label{anomaly}
\theta_{ab} = \sum_i q_i^a q_i^b \theta_0,
\ee
i.e., it is determined entirely by the gauge group representations to which the integrated fermions belong. For the non-Abelian case, the mixed $\theta$-terms (mixed between two gauge groups) vanish identically due to the tracelessness of the representation matrices. The diagonal $\theta$-terms are given by a similar formula with the charges replaced by the trace over the generators of the group in the representation of the partons. A single Dirac fermion in the fundamental representation contributes $\theta=\theta_0$. This calculation is robust against inclusion of interactions\cite{thooft1980} as recently discussed in Ref.~\onlinecite{Mulligan:2010hi}.

From Eq.~\eqref{anomaly} it follows immediately that $N_c$ partons of electric charge $e/N_c$ generate a $\theta$-term for the Maxwell field with $\theta=\theta_0/N_c$, that is $\theta=\pi/N_c$ if the partons realize a $T$-invariant topological insulator. No reference to the emergent gauge group is necessary. In the Abelian case it is advantageous to ensure that no mixed $\theta$-terms involving the Maxwell field and an emergent gauge field are generated. Such mixed terms give rise to extra contributions to the effective $\theta$-term for the Maxwell field (denoted by ``$\mathrm{em}$'') once the emergent gauge fields are integrated out. The latter is a delicate thing to do in the Coulomb phase where the emergent gauge bosons are gapless, but these mixed terms would alter the topological properties of the Maxwell field. Vanishing of the mixed terms requires that for any $a \neq\mathrm{em}$,
\be\label{AbelianAnomaly}
\sum_i q_i^a q_i^\mathrm{em} =0.
\ee
If all partons have the same electromagnetic charge $q_i^\mathrm{em} = 1/N_c$, this reads $\sum_i q_i^a=0$, which is simply the requirement that the electron is gauge-invariant under the emergent gauge fields. This, by construction, is automatically satisfied in our Abelian models presented above. Alternatively, Eq.~(\ref{AbelianAnomaly}) follows from the orthogonality of the generators $\Tr t_at_\mathrm{em}=0$, $a=A,B$ (Sec.~\ref{SecAbelianModels}). On the other hand, the general non-Abelian models introduced in Sec.~\ref{generalmodel} have, for every gauge group factor labeled by $a$, a total of $N_c^a$ partons with emergent charge $q^a=1/N_c^a$ and a Maxwell electric charge that is more or less unconstrained up to the overall condition Eq.~\eqref{chargeconstraint}. In this case mixed terms will be generated. The final value quoted for $\theta$ in Ref.~\onlinecite{Maciejko:2010tx} is only obtained after integrating out the emergent gauge fields in this case.

As an explicit example, take the Abelian $N_c=3$ parton model, model A (Sec.~\ref{SecAbelianModels}). Assume that all partons have a topologically nontrivial mass, i.e., $\theta_0=\pi$. Using the chiral symmetry to rotate the phase of all three mass terms to a real and positive mass, we generate via the ABJ anomaly an electromagnetic $\theta$-term with $ \theta=C \pi$, where
\be
C = \sum_i \left ( q^\mathrm{em}_i \right )^2 = \frac{1}{3}.
\ee
No off-diagonal $F_\mathrm{em} \wedge F_{A}$ or $F_\mathrm{em} \wedge F_{B}$ terms are generated, because the corresponding anomalies
$\sum_i q^\mathrm{em}_i q^A_i$ and $\sum_i q^\mathrm{em}_i q^B_i$ vanish. In the emergent sector, nonzero $\theta$-terms will be generated,
\begin{align}
\label{statistictheta}
\theta_{AA}&= \sum_i (q^A_i)^2 \pi = 6 \pi, \quad \theta_{BB} = \sum (q^B_i)^2 \pi = 6 \pi,\nonumber\\
\theta_{AB}&= \theta_{BA} = q^A_i q^B_i \pi = - 3 \pi.
\end{align}
These extra terms play an important role in the charge quantization considerations below. We note that the above calculations only depend on the fermion content of the theory, and therefore can not distinguish between the Coulomb (model A or B) or Higgs (model C) realizations of this particular fermion content.

\subsection{Topological quantization of the axion angle}

In a topological band insulator, the low-energy description of the electromagnetic response at energy scales below the gap is in terms of a topological field theory with a $S_{\theta}\sim \theta \b{E} \cdot \b{B}$ term.\cite{Qi2008} From the point of view of the low-energy theory, the partition function is $T$-invariant only if $e^{i S_{\theta} }$ (or $e^{-S_\theta}$ in Euclidean signature) is $T$-invariant which, together with the Dirac quantization of fluxes appearing in $S_\theta$, constraints $\theta$ to take discrete values. In the fTI, after integrating out the massive partons, the low-energy electromagnetic response still takes the form $\theta \b{E} \cdot \b{B}$, but with possibly fractional values for $\theta$. The fractional values of $\theta$ allowed in this case follow from modified flux quantization conditions due to the presence of the emergent gauge fields.

\subsubsection{Quantization in the Abelian Coulomb phases}
\label{QuantAbelianModels}

Consider the $\theta$-term for the Maxwell field, $\frac{\theta_\mathrm{eff} e^2}{32\pi^2}\epsilon^{\mu\nu\lambda\rho}F_{\mu\nu}F_{\lambda\rho}$. The quantization condition on $\theta_\mathrm{eff}$ follows directly from the Dirac quantization condition for magnetic charges, and the argument applies either in Euclidean\cite{Witten1995,Levin2009} or Minkowski spacetime\cite{Qi2008,vazifeh2010} (with certain conditions on the electromagnetic fields in the latter case). We consider the Euclidean case for simplicity. With periodic boundary conditions, spacetime is topologically equivalent to the $4$-torus $T^4$. The minimal value of the spacetime integral $S_\theta$ of the $\theta$-term $\frac{i\theta_\mathrm{eff} e^2}{32\pi^2}\epsilon_{\mu\nu\lambda\rho}F_{\mu\nu}F_{\lambda\rho}$ is obtained when the smallest allowed magnetic monopole is inserted inside both $2$-tori $T^2_{12}$ and $T^2_{34}$ where $T^4\cong T^2_{12}\times T^2_{34}$, and $T^2_{\mu\nu}$ is the $2$-torus generated by directions $\mu$ and $\nu$. In Euclidean spacetime, all directions can be taken to be spacelike and thus all fluxes $F_{\mu\nu}$ are magnetic. We obtain
\begin{align}\nonumber
 S_\theta&\equiv\frac{i\theta_\mathrm{eff} e^2}{32\pi^2}\int_{T^4}d^4x\,\epsilon_{\mu\nu\lambda\rho}F_{\mu\nu}F_{\lambda\rho}\nonumber\\
 &=\frac{i\theta_\mathrm{eff} e^2}{4\pi^2}\int_{T^2_{12}}dx_1dx_2\,F_{12}\int_{T^2_{34}}dx_3dx_4\,F_{34},
\end{align}
where the factor of $8$ comes from the permutations of the $\epsilon$-tensor.\cite{NoteSpinMfld} If the fundamental charge is $e$, the smallest allowed magnetic monopole has magnetic flux
\be
\int_{T^2}F=\frac{2\pi}{e} \equiv B_0,
\ee
and $S_\theta=i\theta_\mathrm{eff}$, i.e., $\theta_\mathrm{eff}$ is periodic with period $2\pi$. In the parton model however, the fundamental charge is now $e/N_c$, and the above argument would yield $\int_{T^2}F=\frac{2 \pi N_c}{e}$ and $S_\theta= iN_c^2 \theta_\mathrm{eff}$, which means that $\theta_\mathrm{eff}$ would have period $\frac{2\pi}{N_c^2}$. Also, the requirement that the minimal allowed magnetic Maxwell flux be $N_c B_0$ and no longer $B_0$ seems to be in stark contrast to the real world, where $B_0$ fluxes have certainly been realized. Both of
those puzzles get resolved by taking into account that in models with fractional charges interacting with emergent gauge fields, Maxwell magnetic charges can be accompanied by ``color'' magnetic charges, i.e., magnetic charges of the emergent gauge fields.\cite{thooft1976,preskill1983,topdegeneracy} We now explain how this increases the periodicity of $\theta_\mathrm{eff}$ from $\frac{2\pi}{N_c^2}$ to $\frac{2\pi}{N_c}$. In order to do that, we must first review how the Dirac quantization condition is modified in the presence of multiple Abelian gauge fields.

For multiple Abelian gauge fields, the Dirac quantization condition $q_eq_m\in2\pi\mathbb{Z}$ with $q_e$, $q_m$ the electric and magnetic fluxes, respectively (in the first example, we had $q_e=e$ and $q_m=\frac{2\pi}{e}$), through a closed $2$-manifold such as $T^2$, is replaced by the more generic condition
$\sum_a q_e^aq_m^a \,\in \,2\pi\mathbb{Z}$, where again the superscript $a$ labels the various Abelian gauge group factors, including the Maxwell gauge group. For any given gauge group $a$, $q_e^aq_m^a$ does not have to be an integer multiple of $2 \pi$. Therefore a minimal $B_0$ flux is allowed, even though the product $q_e^aq_m^a$ of the Maxwell magnetic charge producing this flux and the electric charge of a parton would be $2 \pi/N_c$, as long as there are also color magnetic fluxes present. This is possible because $q_e$ for the parton is nonzero for the emergent gauge groups. Outside the topological insulator, the emergent fields are confined and their magnetic flux has no physical consequence. Indeed, confinement of the color electric fluxes corresponds to condensation of the magnetic fluxes,\cite{thooft1978} so the latter fluctuate wildly and there is no energy cost associated with them. Therefore, we recover the fact that in a topologically trivial insulator a $B_0$ flux is possible. Inside a fTI, the emergent $U(1)$ gauge field is deconfined and the color magnetic flux does have an effect.

To determine the periodicity of $\theta$, we need to consider the contribution of all gauge fields to the action $S_\theta$. We consider the general topological term Eq.~(\ref{stheta}) with $\mathcal{M}=T^4$, together with the result of the anomaly calculation Eq.~(\ref{anomaly}). Recall that $q_i^a$ denotes the charge of the $i$th parton under the $a$th gauge group in units of the gauge coupling, so that $q_e^a=e q_i^a$ for the parton. Since $\int_{T^2}F^a=q_m^a$, one finds
\begin{align}
S_\theta &= \frac{i\theta_0}{4 \pi^2} \sum_{a,b,i} \left [ (e q_i^a) (e q^i_b) \right ] \, \left [ q_m^a q_m^b \right ]\nonumber\\
& =\frac{i\theta_0}{4 \pi^2} \sum_{i} \left [ \sum_a (q_e)_i^a  (q_m)_i^a \right ]^2.
\end{align}
The Dirac quantization condition $\sum_a q_e^a \, q_m^a \,\in \,2\pi\mathbb{Z}$ ensures that this is $\theta_0$ times a sum of integers squared, so $\theta_0$ has the standard $2 \pi$ periodicity. The Maxwell $\theta$ angle is $\theta_0/N_c$, according to Eq.~\eqref{anomaly}, and so has periodicity $2 \pi/N_c$ as announced earlier.

We now demonstrate this explicitly on the example of model A. The Dirac quantization condition, i.e., the condition that the parton wave function should be single-valued, allows a $B_0$ flux for the Maxwell field together with a color magnetic flux $B_0/3$ for the (say) $U(1)_B$ magnetic field. To see this, we note that the phase $\alpha_i$ by which the wave function of the $i$ parton changes when taken around a loop enclosing this flux is
\be
\alpha_i = 2 \pi \left(e q_i^B \frac{B_0}{3} + e q_i^\mathrm{em} B_0\right),
\ee
which yields $\alpha_1=0$, $\alpha_2= 2 \pi$, $\alpha_3=0$ and the wave function is indeed single-valued. A calculation of the $\theta_{ab}$ angles in model A, similar to Eq.~\eqref{statistictheta} but with a general $\theta_0$, yields $\theta_\mathrm{em} = \theta_0/3$ and $\theta_{BB}=6 \theta_0$, and we obtain with $a,b\in\{\mathrm{em},A,B\}$ and in Lorentzian signature
\be
e^{iS_\theta} =e^{i \sum_{a,b} \frac{\theta_{ab}}{2 \pi} \frac{e^2}{2 \pi} \int d^4x\,\b{E}_{a} \cdot \b{B}_{b}} =
e^{i \left(6 \theta_0 \times\frac{1}{3^2} + \frac{\theta_0}{3}\times 1^2\right)} = e^{i \theta_0},\label{eq:fluxquant}
\ee
as announced earlier. $\theta_0$ has periodicity $2 \pi$, and hence the Maxwell $\theta$ angle has periodicity $2 \pi/3$.

\subsubsection{Quantization in the non-Abelian Coulomb phases}
\label{quantNAmodels}

In the non-Abelian case, the discussion of charge quantization closely follows the Abelian case discussed above. The proper quantization condition for magnetic fluxes in that case is
$\mathbf{e}_a\cdot\mathbf{m}_b\in2\pi\mathbb{Z}\delta_{ab}$, where $\mathbf{e}_a$ is a ``electric flux vector'', $\mathbf{m}_b$ is a ``magnetic flux vector'', and the indices $a,b$ run over the generators of the gauge group.\cite{topdegeneracy,Englert:1976ng,Goddard:1976qe} Mathematically, $\mathbf{e}_a$ is a vector in the weight lattice of the Lie algebra of the gauge group, and the quantization condition defines $\mathbf{m}_b$ as a vector in the dual weight lattice, i.e., the ``reciprocal'' weight lattice.\cite{Georgi:1982jb}

Let us consider the example of a non-Abelian Coulomb phase with $N_c=3$ partons and $SU(3)$ gauge group (model B). Including electromagnetism, the total gauge group is $SU(3)\times U(1)_\mathrm{em}$ which has the following Cartan generators,
    \begin{align}\label{CartanSU3}
    H_1&=\frac{g}{\sqrt{2}}\lambda_3=\frac{g}{\sqrt{2}}\left(
    \begin{array}{ccc}
    1 & 0 & 0\\
    0 & -1 & 0\\
    0 & 0 & 0
    \end{array}
    \right),\nonumber\\
    H_2&=\frac{g}{\sqrt{2}}\lambda_8=\frac{g}{\sqrt{6}}\left(
    \begin{array}{ccc}
    1 & 0 & 0\\
    0 & 1 & 0\\
    0 & 0 & -2
    \end{array}
    \right),\nonumber\\
    H_3&=\left(
    \begin{array}{ccc}
    \frac{e}{3} & 0 & 0\\
    0 & \frac{e}{3} & 0\\
    0 & 0 & \frac{e}{3}
    \end{array}
    \right),
    \end{align}
where $\lambda_3,\lambda_8$ are Gell-Mann matrices, and we have explicitly written the $SU(3)$ and $U(1)_\mathrm{em}$ gauge couplings $g$ and $e$, respectively. $H_1$ and $H_2$ are the two Cartan generators of $SU(3)$,\cite{Georgi:1982jb} normalized to $\Tr(H_aH_b)=g^2\delta_{ab}$, $a,b=1,2$, and $H_3$ is the generator of $U(1)_\mathrm{em}$ with all three quarks having the same electric charge $e/3$. The weight lattice is generated by the fundamental weights which are
    \begin{align}\label{weights}
    \mathbf{e}_1&=\left(\frac{g}{\sqrt{2}},\frac{g}{\sqrt{6}},\frac{e}{3}\right),\nonumber\\
    \mathbf{e}_2&=\left(-\frac{g}{\sqrt{2}},\frac{g}{\sqrt{6}},\frac{e}{3}\right),\nonumber\\
    \mathbf{e}_3&=\left(0,-\frac{2g}{\sqrt{6}},\frac{e}{3}\right).
    \end{align}
    We recall that $e^i_a$ is the eigenvalue of the Cartan generator $H_i$ associated with the $a$th common eigenvector $|\mathbf{e}_a\rangle$ of all three Cartan generators, i.e., $H_i|\mathbf{e}_a\rangle=e^i_a|\mathbf{e}_a\rangle$.\cite{Georgi:1982jb} The first two entries of $\mathbf{e}_a$ correspond to non-Abelian $SU(3)$ ``color'' charges, and the last entry corresponds to the usual $U(1)_\mathrm{em}$ electric charge. The dual weight lattice is generated by its own fundamental weights, which are defined as the ``reciprocal lattice basis vectors'' $\mathbf{m}_b$,
    \begin{align}\label{quant}
    \mathbf{e}_a\cdot\mathbf{m}_b=2\pi\delta_{ab},\,a,b=1,2,3.
    \end{align}
    The linear system Eq.~(\ref{weights}), (\ref{quant}) is easily solved to yield\cite{noteweights}
    \begin{align}\label{dualweights}
    \mathbf{m}_1&=2\pi\left(\frac{1}{\sqrt{2}g},\frac{1}{\sqrt{6}g},\frac{1}{e}\right),\nonumber\\
    \mathbf{m}_2&=2\pi\left(-\frac{1}{\sqrt{2}g},\frac{1}{\sqrt{6}g},\frac{1}{e}\right),\nonumber\\
    \mathbf{m}_3&=2\pi\left(0,-\frac{2}{\sqrt{6}g},\frac{1}{e}\right).
    \end{align}
    The allowed magnetic monopoles, i.e., the allowed magnetic flux configurations, are given by linear combinations of the fundamental dual weights $\mathbf{m}_b$ with integer coefficients,
    \begin{align}\label{duallattice}
    \mathbf{m}=n_1\mathbf{m}_1+n_2\mathbf{m}_2+n_3\mathbf{m}_3,\,n_1,n_2,n_3\in\mathbb{Z}.
    \end{align}
    We are now in a position to discuss the periodicity of $\theta_\mathrm{eff}$. The ``colorless'' magnetic monopole configuration discussed earlier, which led to a periodicity of $\frac{2\pi}{N_c^2}=\frac{2\pi}{9}$, corresponds to the dual weight vector
    \begin{align}\nonumber
    \mathbf{m}=\mathbf{m}_1+\mathbf{m}_2+\mathbf{m}_3=\left(0,0,\frac{6\pi}{e}\right),
    \end{align}
    i.e., $n_1=n_2=n_3=1$ in Eq.~(\ref{duallattice}). However, we now see that this is not the ``smallest'' magnetic monopole: we can choose a smaller monopole for which some of the $n_i$, $i=1,2,3$ are zero. In particular, the smallest monopoles have only one $n_i$ equal to $1$. But as seen in Eq.~(\ref{dualweights}), these monopoles will necessarily be ``colored'', i.e., they will carry some amount of non-Abelian magnetic charge. Let us now evaluate $S_\theta$ for a colored magnetic monopole, say $\mathbf{m}=\mathbf{m}_1$, and see how it affects the periodicity of $\theta_\mathrm{eff}$. Since the monopole is colored, we cannot simply discard the $SU(3)$ $\theta$-term in the effective action [Eq.~(6) in Ref.~\onlinecite{Maciejko:2010tx}]. Denoting by $F$ and $f$ the $U(1)_\mathrm{em}$ and $SU(3)$ field strengths, respectively, we have
    \begin{align}
    S_\theta&=\frac{i\theta_\mathrm{eff}e^2}{32\pi^2}\int_{T^4}d^4x\,\epsilon_{\mu\nu\lambda\rho}F_{\mu\nu}F_{\lambda\rho}\nonumber\\
    &\hspace{10mm}+\frac{i\theta g^2}{32\pi^2}\int_{T^4}d^4x\,\epsilon_{\mu\nu\lambda\rho}f^a_{\mu\nu}f^a_{\lambda\rho}\nonumber\\
    &=\frac{i\theta}{3}\frac{e^2}{4\pi^2}\left(\frac{2\pi}{e}\right)^2+\frac{i\theta g^2}{4\pi^2}
    \left[\left(\frac{2\pi}{\sqrt{2}g}\right)^2+\left(\frac{2\pi}{\sqrt{6}g}\right)^2\right]\nonumber\\
    &=\frac{i\theta}{3}+\frac{i\theta}{2}+\frac{i\theta}{6}=i\theta,\nonumber
    \end{align}
    i.e., $\theta$ has periodicity $2\pi$, which means that $\theta_\mathrm{eff}=\theta/3$ has periodicity $\frac{2\pi}{3}$, as claimed in Ref.~\onlinecite{Maciejko:2010tx}. It is readily checked that the other ``minimal'' monopoles, $\mathbf{m}=\mathbf{m}_2$ and $\mathbf{m}=\mathbf{m}_3$, give the same quantization condition.

\subsubsection{Quantization in the Higgs phases}

The analysis of charge quantization in the Higgs models is very similar to the Abelian case. The basic statement is as before. A Maxwell $B_0$ flux is possible only if it is accompanied by color magnetic flux of the unbroken gauge group $H$, otherwise only $N_c B_0$ is possible. While color magnetic flux before was quantized in integer multiples of a basic unit, it only now takes a finite number of discrete values. For example, in our model C, magnetic flux for $H=\mathbb{Z}_3$ can only take three different values: three units of color magnetic flux are equivalent to no color magnetic flux.

\section{Ground state degeneracy}
\label{SECGSD}

\subsection{Ground state degeneracy on $T^3$}
\label{GSDonT3}

Recently it has been shown\cite{Swingle:2010rf} that a fractional $\theta$ angle in a $T$-invariant, gapped system necessarily implies multiple ground states on $T^3$. However, a deconfined gauge theory may or may not be gapped, and a unique ground state on $T^3$ is possible if the system is gapless. In particular, we will show that the Abelian Coulomb phases contain gapless but electromagnetically neutral gauge bosons, and have a unique ground state on $T^3$. On the other hand, the non-Abelian Coulomb phases do exhibit a nontrivial ground-state degeneracy on $T^3$, even though they avoid the theorem of Ref.~\onlinecite{Swingle:2010rf} due to the extra massless degrees of freedom. In the Higgs models, the theory obeys all the assumptions of the theorem and there has to be multiple ground states on $T^3$. We will see that this is indeed the case. Therefore, various realizations of a fTI with the same value of the Maxwell $\theta$ angle can be distinguished by their topological ground state degeneracy on $T^3$. This is similar to the fact that the Hall conductance, although a topological invariant, is not sufficient to fully characterize the topological order in a FQH system.\cite{topdegeneracy}

Before we discuss our three different models in detail, we must first define precisely what we mean by ground state degeneracy. As discussed in detail in the previous sections, in at least two of our three models (the Coulomb models A and B), the theory contains gapless degrees of freedom. In order to arrive at a meaningful definition of topological ground state degeneracy in a gapless system, we need to study the theory for a finite-size system. Denoting by $L$ the linear system size, the massless degrees of freedom will develop a finite-size gap of order $1/L$. In order to focus on the ground state manifold, we want to study the theory at energies below that finite-size gap. For a finite-size system, the ground state degeneracy will be generally lifted by nonperturbative effects\cite{topdegeneracy} corresponding to the tunneling of fractionally charged partons around noncontractible loops in $T^3$, which leads to an energy splitting of order $\sim e^{-m_\mathrm{gap} L}/L$ where $m_\mathrm{gap}$ is the dynamically generated parton mass gap already present at infinite volume. For finite $L$, these states are truly degenerate only in the strict $m_\mathrm{gap} \rightarrow \infty$ limit (i.e., for infinitely massive partons), but even at finite $m_\mathrm{gap}$ we can identify the ground states by a finite-size scaling analysis of the many-body spectrum.

\subsubsection{Abelian models}
\label{SecGSDT3Abelian}

We first investigate the question of ground state degeneracy on $T^3$ in the simplest Abelian Coulomb phase, model A. When studying the model on $T^3$, the ground state is unique. As the emergent $U(1)$ fields are noninteracting at low energies, this is simply a question of quantizing Maxwell electrodynamics. As is well known, $U(1)^{N_c-1}$ gauge theory on $T^3$ in the Coulomb phase is simply equivalent to a set of $2(N_c-1)$ decoupled three-dimensional (in the $x,y,z$ directions on $T^3$) harmonic oscillators, i.e., a harmonic oscillator for each of the $U(1)$ gauge fields with two polarizations. For a finite size system, the ground state is unique and corresponds to each of these harmonic oscillators being in their ground state. The gap to the first excited state is of order $\sim 1/L$ where $L$ is the linear system size, but vanishes in the $L\rightarrow\infty$ limit. According to Ref.~\onlinecite{Swingle:2010rf}, a gapped system can not have a fractional $\theta$ angle if it has a unique ground state on $T^3$. As mentioned, the way our simple Abelian model avoids this constraint is because it is not gapped (for infinite size). To see how the extra free photons can circumvent the no-go theorem of Ref.~\onlinecite{Swingle:2010rf}, we observe that the argument of Ref.~\onlinecite{Swingle:2010rf} is essentially a Lorentzian version of the usual Dirac quantization argument. Consider a gapped scenario where a $2$-torus $T^2$ of the spatial $T^3$ is pierced by a magnetic flux $B_z$. In the presence of a unit flux $B_z = B_0=\frac{2 \pi}{e}$, the ground-state to ground-state (G2G) amplitude when inserting the same minimal flux through the noncontractible loop in the $z$ direction picks up a phase given by
\be e^{iS_{\theta}} = e^{i \frac{\theta}{2 \pi} \frac{e^2}{2 \pi} \int d^4x\,\b{E}_\mathrm{em} \cdot \b{B}_\mathrm{em}} =
e^{i \theta}. \ee The time-reversed process picks up a phase $e^{- i \theta}$, therefore for a $T$-invariant theory we require $e^{i \theta} = e^{- i \theta}$ and $\theta$ has to be an integer multiple of $\pi$.

This argument relies on the fact that the minimal magnetic flux is $B_0$. In a theory of a Maxwell gauge field alone with partons of charge $e/3$ a magnetic flux $B_0$ is not consistent with single-valuedness of the parton wave function; rather, the minimal flux allowed is $3 B_0$. In this case, the above G2G amplitude picks up a phase factor of $e^{i9 \theta}$, hence apparently any multiple of $\pi/9$ would be an allowed $T$-invariant value for $\theta$. Clearly, this situation can not correspond to nature as we know it since a $B_0$ flux can exist. At this stage, the discussion is completely parallel to that of Dirac quantization in Sec.~\ref{QuantAbelianModels}. As we have seen in Eq.~(\ref{eq:fluxquant}), a basic $B_0$ flux is allowed as long as it is accompanied by a magnetic flux of the emergent gauge fields. The Dirac quantization condition implies that the allowed combinations of fluxes are exactly the ones for which $e^{i S}$ evaluates to $e^{i n \pi}$ for integer $n$. The G2G amplitude is $T$-invariant despite the fact that the Maxwell $\theta$ angle takes the fractional value $\pi/3$. In fact, Dirac quantization will ensure that the same happens in all the Abelian Coulomb models proposed in Ref.~\onlinecite{Maciejko:2010tx}. The no-go theorem is avoided even without ground state degeneracy on $T^3$ due to the presence of extra massless gauge fields.

\subsubsection{Non-Abelian models}
\label{GSDNonAbelian}

The Coulomb phase of $SU(N_c)$ gauge theory on $T^3$ has $N_c^3$ degenerate ground states.\cite{Sato:2007xc} There are two complementary ways of establishing this result, one by keeping the fundamental partons in the spectrum, and the other by sending the parton mass to infinity, $m_\mathrm{gap}\rightarrow\infty$, and considering the pure $SU(N_c)$ gauge theory. For concreteness we will discuss the case $N_c=3$. In the presence of fundamental quarks, one can construct a topological symmetry algebra consisting of operators $U_a$, $a=1,2,3$ which insert a $2\pi$ flux of the $U(1)_\mathrm{em}$ gauge field through the $a$th noncontractible loop of $T^3$, and operators $T_a$ which move a parton around the $a$th noncontractible loop. In the $T$-invariant case, we have $[T_a,T_b]=[U_a,U_b]=0$, $a,b=1,2,3$, but\cite{Sato:2007xc}
\begin{align}\label{TSA}
T_aU_b=e^{-2\pi i/3\delta_{ab}}U_bT_a,
\end{align}
which simply means that partons can pick up a nontrivial $U(1)_\mathrm{em}$ Aharonov-Bohm phase because they are fractionally charged. The operator $T_a$ moves a parton around a closed loop and is clearly a symmetry of the partition function. The operator $U_a$ is also a symmetry of the partition function because the $U(1)_\mathrm{em}$ phase $e^{2\pi i/3}$ acquired by the partons is also an element of $\mathbb{Z}_3=\{1,e^{2\pi i/3},e^{4\pi i/3}\}$, the center of $SU(3)$, and can thus be gauged away. The existence of an algebra of operators which commute with the Hamiltonian but not among themselves implies the degeneracy of the energy eigenstates. The symmetry algebra Eq.~(\ref{TSA}) is somewhat reminiscent of the topological symmetry algebra in the FQH states.\cite{topdegeneracy} A representation of the algebra Eq.~(\ref{TSA}) can be constructed by first constructing a representation of the Abelian subalgebra generated by the $T_a$, i.e., by taking a set of states which diagonalizes the $T_a$ simultaneously, $T_a|\boldsymbol{\eta}\rangle=e^{i\eta_a}|\boldsymbol{\eta}\rangle$ with $\eta=(\eta_1,\eta_2,\eta_3)$. We first pick a particular noncontractible loop $a$. By applying $U_a$ to $|\boldsymbol{\eta}\rangle$ repeatedly and using the commutation relations (\ref{TSA}), one can show that $\{|\boldsymbol{\eta}\rangle=U_a^3|\boldsymbol{\eta}\rangle,U_a|\boldsymbol{\eta}\rangle, U_a^2|\boldsymbol{\eta}\rangle\}$ is a $3$-dimensional representation of the subalgebra generated by $T_1,T_2,T_3$ and the $U_a$ corresponding to this specific loop. For general $N_c$, we would obtain the $N_c$-dimensional representation $\{|\boldsymbol{\eta}\rangle=U_a^{N_c}|\boldsymbol{\eta}\rangle,U_a|\boldsymbol{\eta}\rangle, U_a^2|\boldsymbol{\eta}\rangle,\ldots,U_a^{N_c-1}|\boldsymbol{\eta}\rangle\}$. The analysis can then be repeated by starting from this enlarged set of states and applying the $U_a$ corresponding to the remaining noncontractible loops. At the end of this process one finds that the dimension of the representation of the full symmetry algebra (\ref{TSA}), which is the same as the ground state degeneracy, is $N_c^{b_1(M)}$ where $b_1(M)=\dim H_1(M,\mathbb{R})$, the first Betti number\cite{nakahara} of the spatial manifold $M$, corresponds to the number of noncontractible loops in $M$. It is the dimension of the first homology group $H_1(M,\mathbb{R})$ of $M$ with real coefficients. For $T^3$ we have $b_1(M)=3$.

\begin{figure}[t]
\begin{center}
\includegraphics[width=3in]{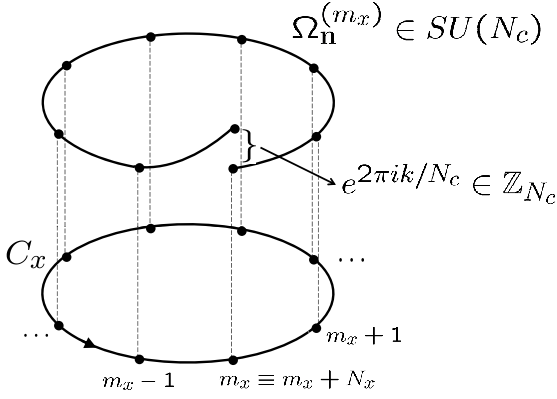}
\end{center}
\caption{$SU(N_c)$ gauge transformation $\Omega$ on a noncontractible loop $C_x$ in the spatial manifold [Eq.~(\ref{GaugeTransZ})], periodic up to an element $e^{2\pi ik/N_c}$ of the center $\mathbb{Z}_{N_c}$ of $SU(N_c)$, and under which the global Wilson loop $W(C_x)$ transforms nontrivially as $W(C_x)\rightarrow e^{-2\pi ik/N_c}W(C_x)$.}
\label{fig:tmatrix}
\end{figure}

Alternatively, this result can be obtained by studying the pure $SU(N_c)$ gauge theory which is the low-energy effective description at energies much less than the parton mass gap $m_\mathrm{gap}$. In this language, the deconfined phase on $T^3$ corresponds to the condensation of spatial Wilson loops, i.e., a spatial version of the condensation of the Polyakov loop\cite{polyakov1978} (temporal Wilson loop). If all fields are in the adjoint (such as is the case once the partons have been integrated out), $SU(N_c)$ gauge theory formulated on a spacetime 4-manifold $\mathcal{M}$ with nontrivial first homology group $H_1(\mathcal{M})\neq 0$ develops a global $\mathbb{Z}_{N_c}$ symmetry\cite{Witten:1998zw,Aharony:1998qu} originating from the center of $SU(N_c)$. This global symmetry is generated by large gauge transformations, where the gauge parameter is not periodic on $T^3$ but only periodic up to an element of the center. An order parameter for spontaneous breaking of this center symmetry is a spatial Wilson loop along a noncontractible loop $C_a$,
\begin{align}\label{GlobalWL}
W(C_a) = \Tr P\exp\left(ig \oint_{C_a} a\right),
\end{align}
where $g$ is the Yang-Mills gauge coupling, $a$ is the emergent $SU(N_c)$ gauge potential, $P$ indicates path-ordering along the loop, and the trace is in the fundamental representation. The fact that $W(C_a)$ transforms nontrivially under the center of $SU(N_c)$ is easily seen by regularizing the theory on a lattice. Denote lattice sites by a triplet of integers $\b{n}=(n_x,n_y,n_z)$, with $n_a=1,\ldots,N_a$, $a=x,y,z$, and periodic boundary conditions on the link variables $U_{\b{n},\mu}=U_{\b{n}+N_a\hat{a},\mu}$, $\mu=t,x,y,z$, in all spatial directions $a$. Consider the following family of $\mathbb{Z}_{N_c}\subset SU(N_c)$ local gauge transformations $\Omega_\b{n}^{(m_x)}$ parameterized by an integer $m_x$,
\begin{align}\label{GaugeTransZ}
\Omega_\b{n}^{(m_x)}=\left\{
\begin{array}{cc}
1, & n_x=1,\ldots,m_x,\ldots,N_x,\\
e^{2\pi ik/N_c}, & n_x=m_x+N_x,
\end{array}
\right.
\end{align}
with $k=1,\ldots,N_c$. Although the gauge transformation Eq.~(\ref{GaugeTransZ}) is not periodic $\Omega_\b{n}^{(m_x)}\neq\Omega_{\b{n}+N_x\hat{x}}^{(m_x)}$, it still is a valid gauge transformation because the usual gauge-invariant operators $\sum_\square\Tr U_\square$ are invariant under \emph{any} local gauge transformation, including multivalued ones (recall that the plaquette variable $U_\square\equiv U_{\b{n},\mu}U_{\b{n}+\hat{\mu},\nu}U^\dag_{\b{n}+\hat{\nu},\mu}U^\dag_{\b{n},\nu}$ transforms in the adjoint as $U_\square\xrightarrow{\Omega}\Omega_\b{n}U_\square\Omega_\b{n}^\dag$ for an arbitrary local gauge transformation $\Omega_\b{n}$). Another way to say this is that although $\Omega_\b{n}^{(m_x)}$ is not periodic as a $SU(N_c)$ gauge transformation, it is periodic as a $SU(N_c)/\mathbb{Z}_{N_c}$ gauge transformation. Since the gauge fields transform trivially under $\mathbb{Z}_{N_c}$ because they are in the adjoint, such a gauge transformation preserves the periodic boundary conditions on the gauge fields.

We can now show that Eq.~(\ref{GlobalWL}) transforms nontrivially under $\Omega_\b{n}^{(m_x)}$. The global Wilson loop Eq.~(\ref{GlobalWL}) has a natural lattice regularization,
\begin{align}
W(C_a)=\Tr\prod_{\b{n}\in C_a}U_{\b{n},a}.
\end{align}
Denoting $\b{n}_\perp\equiv(n_y,n_z)$, we see that $W(C_x)$ transforms under $\Omega^{(m_x)}_\b{n}$ as
\begin{align}\label{WCtransform}
W(C_x)&=\Tr U_{(1,\b{n}_\perp),x}U_{(2,\b{n}_\perp),x}\cdots U_{(N_x,\b{n}_\perp),x}\nonumber\\
&=\Tr U_{(m_x,\b{n}_\perp),x}U_{(m_x+1,\b{n}_\perp),x}\cdots U_{(m_x+N_x-1,\b{n}_\perp)}\nonumber\\
&\xrightarrow{\Omega}\Tr(1\cdot U_{(m_x,\b{n}_\perp),x}\cdot 1)(1\cdot U_{(m_x+1,\b{n}_\perp),x}\cdot 1)\cdots\nonumber\\
&\hspace{5mm}\times(1\cdot U_{(m_x+N_x-1,\b{n}_\perp)}\cdot e^{-2\pi ik/N_c})\nonumber\\
&=e^{-2\pi ik/N_c}W(C_x),
\end{align}
using the periodicity of the trace to shift the base point of the loop from $(1,\b{n}_\perp)$ to $(m_x,\b{n}_\perp)$ and the fact that the link variables $U_{\b{n},\mu}$ transform as $U_{\b{n},\mu}\xrightarrow{\Omega}\Omega_\b{n}U_{\b{n},\mu}\Omega^\dag_{\b{n}+\hat{\mu}}$. When the theory is quantized, expectation values of operators such as $W(C_a)$ are computed by averaging over all $SU(N_c)$ gauge transformations that are periodic $\Omega_\b{n}=\Omega_{\b{n}+N_a\hat{a}}$. Note that the transformation law Eq.~(\ref{WCtransform}) is independent of the local data $m_x$ specifying where the discontinuity of the gauge function $\Omega^{(m_x)}_\b{n}$ occurs. Therefore, although such local data can be lost by performing the $SU(N_c)$ gauge averaging, the global data (i.e., the parameter $k$) is not averaged out and $W(C_a)$ can develop a nonzero expectation value. This is why the transformation law Eq.~(\ref{WCtransform}) is in fact a global symmetry. As a result, spontaneous breaking of this symmetry $\langle W(C_a)\rangle\neq 0$ does not violate Elitzur's theorem.\cite{elitzur1975}

The Wilson loop Eq.~(\ref{GlobalWL}) can be interpreted as the semiclassical process of creating a heavy parton-antiparton pair at $x=0$ and annihilating them again at $x=L/2$ along the noncontractible loop $C_a$, or in other words the worldline of a virtual parton taken around the loop once. Therefore, we recover essentially the same physics as when we explicitly considered the fundamental partons in the previous approach of establishing the ground state degeneracy. The Polyakov loop is the finite-temperature version of Eq.~(\ref{GlobalWL}), where the spatial manifold need not have a nontrivial first homology but the relevant noncontractible loop in that case is the periodic imaginary time direction $\tau\in[0,\beta]$, with $\beta=1/T$ the inverse temperature. In this case, the logarithm of the Polyakov loop is the negative of the free energy associated with a single parton. In a confined phase, this free energy is infinite as it takes an infinite amount of energy to separate a single color charge from its color charge conjugate, and the Polyakov loop has zero expectation value. Center symmetry is unbroken. On the other hand, in a deconfined phase the Polyakov loop typically has a nonzero expectation value and center symmetry is spontaneously broken. In our case we are considering a zero-temperature theory and the spatial Wilson loop Eq.~(\ref{GlobalWL}) characterizes G2G amplitudes at zero temperature. As each $W(C_a)$ carries charge under $\mathbb{Z}_{N_c}$, there are $N_c$ different values that $W(C_a)$ can take for each noncontractible loop $C_a$. Therefore, there will be $N_c^3$ degenerate ground states on $T^3$ and $N_c^{b_1(M)}$ ground states for a general spatial manifold $M$ without boundary.

We remark that this ground state degeneracy was not required by the theorem of Ref.~\onlinecite{Swingle:2010rf}. Whether this $N_c^{b_1(M)}$ degeneracy actually occurs depends on details of the system beyond the number of partons and the gauge group. In particular, if the extra gapless neutral matter added in order to drive the non-Abelian emergent gauge field into a Coulomb phase is in the {\it fundamental} representation of the gauge group, there is no global center symmetry and we expect that the ground state would be unique. Another example where spontaneous breaking of the global $\mathbb{Z}_{N_c}$ symmetry (and hence presumably the corresponding ground state degeneracy) does not necessarily occur is $\mathcal{N}=4$ SYM theory. As long as one imposes periodic boundary conditions for the adjoint fermions along all noncontractible loops, the potential for the corresponding Wilson loop is flat due to supersymmetry. The ground state expectation value of the Wilson loop is one more modulus in the theory. One can tune its expectation value arbitrarily, hence we can obtain a deconfined phase with $\mathbb{Z}_{N_c}$ either broken or unbroken.

\subsubsection{Higgs models}

Because the Higgs models of Sec.~\ref{HiggsModels} are completely gapped, they must have multiple ground states on $T^3$ to be consistent with the theorem of Ref.~\onlinecite{Swingle:2010rf}. As we necessarily have unbroken discrete gauge groups, this is indeed ensured. Discrete gauge groups give rise to degenerate ground states that differ by the value of the Wilson loop of the discrete gauge field around noncontractible loops, i.e., discrete global fluxes. For example, in our model C with its $\mathbb{Z}_3$ unbroken gauge group, there are three different discrete fluxes per noncontractible loop of the spatial manifold $T^3$, which label the various ground states for a total ground state degeneracy of $3^3=27$. A generic $SU(N_c)$ Higgs model on a boundaryless spatial manifold $M$ with all fields in the adjoint, and where $SU(N_c)$ is spontaneously broken to its center $\mathbb{Z}_{N_c}$ will have a ground state degeneracy of $N_c^{b_1(M)}$. This is the three-dimensional version of the $N^2$ ground state degeneracy on $T^2$ in the deconfined phase of $\mathbb{Z}_N$ gauge theory in $2+1$ dimensions.\cite{barkeshli2010b} A similar mechanism is also responsible for the ground state degeneracy in the $\mathbb{Z}_2$ spin liquid model put forward in Ref.~\onlinecite{Swingle:2010rf} as an explicit realization of a fully gapped fTI.

\subsection{Ground state degeneracy on $3$-manifolds with boundaries}

On a $3$-manifold with boundaries, such as the cartesian product $M=\Sigma\times I$ of a Riemann surface $\Sigma$ and an interval $I$ discussed in our previous work,\cite{Maciejko:2010tx} we need to also consider the effect of the Chern-Simons (CS) terms induced on the boundary. To be specific, let us focus on the $M=\Sigma\times I$ case with boundary $\partial M=\Sigma\cup\Sigma$. The nontrivial $\theta$-term of the emergent gauge fields in the bulk induces a CS term on the boundary due to the axion domain wall\cite{Qi2008} between $\theta\neq 0$ inside the fTI and $\theta=0$ outside the fTI. For the case of $\theta=\pi$, the corresponding CS term has level $1/2$. The fact that this term is not gauge-invariant as a purely $(2+1)$-dimensional theory does not matter, because the gauge noninvariance of the boundary theory is compensated by the bulk. This is made clear by writing the $\theta$-term as a manifestly gauge-invariant $\Tr\epsilon^{\mu\nu\lambda\rho}f_{\mu\nu}f_{\lambda\rho}$ term,\cite{Gaiotto:2008sd,Chen2010} where $f$ is the emergent field strength. The emergent gauge field also has a kinetic (Yang-Mills) term $\sim-\frac{1}{g^2}\Tr f_{\mu\nu}f^{\mu\nu}$. (In this section we work with rescaled fields $a_\mu\rightarrow\frac{1}{g}a_\mu$.) %JM: I added the previous sentence to fix the g's.
The details of the ground state degeneracy will depend on the low-energy dynamics of the gauge fields. We have already shown that the gauge fields need to be in a deconfined phase to have a fTI. In the following, we will further assume that the deconfined phase occurs at weak coupling $g\ll 1$, so that we can consider the gauge fields as being essentially free. The situation is less clear in the case of a deconfined but strongly coupled theory such as $\mathcal{N}=4$ SYM theory. %JM: I added the following clarifications for the non-Abelian case.
Just as in the Abelian case, the Yang-Mills Hamiltonian is minimized by setting the emergent electric and magnetic fields to zero, which corresponds to a zero emergent field strength $f_{\mu\nu}=0$. Therefore, the ground state eigenspace is obtained by quantizing the space of flat connections $a_\mu=U\partial_\mu U^{-1}$ with $U:M\times\mathbb{R}\rightarrow SU(N_c)$ a gauge function which is not necessarily periodic on $M$ to allow for global fluxes through noncontractible cycles, and $\mathbb{R}$ denotes time. Denoting by $\omega_\mathrm{CS}=\Tr\left(a\wedge\mathrm{d}a+i\frac{2}{3}a\wedge a\wedge a\right)$ the CS form and using the fact\cite{nakahara} that $\Tr f\wedge f=\mathrm{d}\omega_\mathrm{CS}$ with $f=\mathrm{d}a+ia\wedge a$ the emergent field strength, we have
%XLnote: I think there is no g in the definition of f and \omega_CS here, in the convention we choose. Please doublecheck. %JM: fixed the g's.
\begin{align}\label{FFboundaryCS}
\frac{1}{8\pi^2}\int_{M\times\mathbb{R}}\theta(z)\Tr f\wedge f&=-\frac{1}{8\pi^2}\left(\Delta\theta_{z=-\frac{L}{2}}\int_{\Sigma\times\mathbb{R}}\omega_\mathrm{CS}^+\nonumber\right.\\+
&\left.\Delta\theta_{z=\frac{L}{2}}\int_{\Sigma\times\mathbb{R}}\omega_\mathrm{CS}^-\right),
\end{align}
with $\Delta\theta_{z=\pm\frac{L}{2}}\equiv\theta(z=\pm L/2+\epsilon)-\theta(z=\pm L/2-\epsilon)$, for $\epsilon$ a positive infinitesimal, and $\omega_\mathrm{CS}^\pm$ denotes the CS form evaluated at $a_\mu^\pm\equiv a_\mu(z=\pm L/2)$. Since $z=\pm L/2$ correspond to the boundary between a fTI and the vacuum which is a trivial insulator, we generically have $\Delta\theta_{z=\pm\frac{L}{2}}=(2k_\pm+1)\pi$ where $k_+,k_-\in\mathbb{Z}$. It appears that we have two distinct CS terms, one for each boundary gauge field $a_\mu^\pm$. However, those two CS terms are not independent because the two $(2+1)$-dimensional boundary gauge fields originate from the same $(3+1)$-dimensional gauge function $U$. As a result, $a_\mu^+$ and $a_\mu^-$ are equal up to a gauge transformation of even winding number which leaves the partition function invariant (see Eq.~(\ref{evenwindingnumber}) of Appendix A). The resulting theory is that of a single CS gauge field on $\Sigma\times\mathbb{R}$ with integer level $k\equiv k_++k_-+1$. This CS theory gives a additional contribution to the ground state degeneracy.

In the Abelian case, we obtain a sum of $U(1)_k$ CS terms on the boundary (see Appendix B). However, the ground state degeneracy is not simply the product of the ground state degeneracy for each $U(1)_k$ CS term, because large gauge transformations which mix several $U(1)$ factors give additional constraints on the ground state Hilbert space.\cite{topdegeneracy} For the Abelian $U(1)^2$ theory (Sec.~\ref{SecAbelianModels}) corresponding to $N_c=3$, the ground state degeneracy is $\frac{1}{2}(k+1)(k+2)$ for $\Sigma=T^2$,\cite{topdegeneracy} which reproduces the familiar threefold degeneracy of the $\nu=1/3$ FQH state on the torus for $k=1$. For the $\mathbb{Z}_{N_c}$ Higgs models, the Maxwell term of the $\mathbb{Z}_{N_c}$ gauge theory itself defines a topological field theory and the ground state degeneracy is $N_c^{2\mathfrak{g}}$ where $\mathfrak{g}$ is the genus of $\Sigma$.\cite{WenPartons}

In the non-Abelian case, we obtain a non-Abelian CS term on $\Sigma$.\cite{Maciejko:2010tx}  The ground state degeneracy for a $SU(N_c)_k$ CS theory on $\Sigma$ is equal to the number of conformal blocks of the level-$k$ $SU(N_c)$ Wess-Zumino-Witten (WZW) conformal field theory.\cite{Witten:1988hf} This number has been determined for any gauge group $G$, level $k$ and genus $\mathfrak{g}$ of $\Sigma$,\cite{Verlinde:1988sn,Dijkgraaf:1988tf} but the answer is particularly easy for the special case of the torus, $\mathfrak{g}=1$. In this case the groundstate degeneracy at level $k$ is given by $1/(N_c-1)! \prod_{j=1}^{N_c-1} (k+j) $. For the special case of $k=1$ we are mostly interested in this simply evaluates to $N_c$. For $N_c=3$, the ground state degeneracy is $\frac{1}{2}(k+1)(k+2)$ in complete agreement with the answer found in Ref.~\onlinecite{topdegeneracy}. This is the same as the Abelian $U(1)^2$ theory.

%XLnote: I added the discussion below for more generic manifolds. We cannot have a formula but I think the procedure is well-defined. Please see if you agree.
%JM: I have kept the general discussion but deleted the S\S' example.

This discussion can be generalized to more generic manifolds $M$ which are not of the form $\Sigma\times I$. Due to similar arguments as above, the ground state degeneracy should only depend on the ``zero modes" with vanishing spatial gauge curvature. For each topologically nontrivial loop $C_i$ on the surface $\partial M$, we define a holonomy\cite{holonomyvsWL} or Wilson loop operator $\Gamma(C_i)\equiv P\exp\left(i\oint_{C_i}a\right)$. It is important to note that $C_i$'s are considered equivalent if they can be deformed to each other through some deformations in $M$ (rather than being restricted to $\partial M$). For example, in the case of $T^2\times I$ the four cycles on the two boundary tori are identified pairwise, leading to two independent nontrivial loops. After taking this condition into account, the operators $\Gamma(C_i)$ on distinct topologically nontrivial loops have commutation relations determined by the boundary CS theory in the same way as the $T^2\times I$ case discussed above. In practice this is a difficult problem to solve, and we have not been able to obtain a general formula which accounts for both bulk and boundary degeneracies. However, some simple examples can be studied explicitly. The simplest example of manifold not of the form $M=\Sigma\times I$ is the solid torus $S^1\times D^2$ with $D^2$ the two-dimensional disc. The unique boundary is the torus $T^2$. Because the vacuum has $\theta=0$ and the bulk of the solid torus has $\theta=\pi$ mod $2\pi$, the CS term on the boundary has a level which is an odd multiple of $1/2$. Let us first consider the Abelian case for simplicity, where we have a sum of $U(1)$ CS terms with half-odd-integer level, and focus on a single CS term. The ground state degeneracy in this case can be obtained by adding to the Lagrangian a small Maxwell term with a coefficient which is sent to zero at the end of the calculation.\cite{wen1989b} The quantum dynamics of the global Wilson loops is equivalent to a unit charged particle moving on $T^2$ in the presence of a uniform magnetic field, where the total number of flux quanta passing through $T^2$ is given by the CS level. In the usual FQHE on $T^2$, the CS level $k$ is integer and the number of degenerate states in the lowest Landau level of this effective single-particle quantum mechanics problem\cite{haldane1985} is also $k$. In our situation, the level is a half-odd-integer. This means that the magnetic translation operators $t(\b{L}_x)$ and $t(\b{L}_y)$, where $\b{L}_x$ and $\b{L}_y$ are the two generators of $T^2$, do not commute. Because the effective quantum-mechanical particle lives on $T^2$, we should apply periodic boundary conditions on the single-particle wave function $\psi(x,y)$,
\begin{align}
t(\b{L}_x)\psi=e^{i\phi_x}\psi,
\hspace{5mm}
t(\b{L}_y)\psi=e^{i\phi_y}\psi,\nonumber
\end{align}
where $\phi_x$ and $\phi_y$ are real phases, which implies $[t(\b{L}_x),t(\b{L}_y)]\psi=0$. However, since $[t(\b{L}_x),t(\b{L}_y)]\neq 0$ for a noninteger total number of flux quanta through $T^2$, the only solution is $\psi=0$. This argument holds for each CS term. We expect the same to be true for the non-Abelian fTI. Indeed, for $SU(N_c)$ CS theory on $T^2$ the global Wilson loops should form a representation of the fundamental group $\pi_1(T^2)$ which is Abelian. Therefore these Wilson loops should belong to the maximal Abelian subgroup $U(1)^{N_c-1}$ of $SU(N_c)$,\cite{topdegeneracy,Witten:1988hf} and their dynamics is given by $N_c-1$ quantum-mechanical particles on $T^2$. If the CS level is half-odd-integer as in the present situation, there is again a single ground state $\psi=0$. Although $SU(N_c)$ and $U(1)^{N_c-1}$ CS theories are not equivalent due to the Weyl group of $SU(N_c)$, in both cases this trivial ground state is the only ground state which is consistent with periodic boundary conditions in both directions. Therefore, the boundary CS theory does not contribute to the ground state degeneracy for the solid torus. The only contribution to the ground state degeneracy comes from the bulk. For an Abelian fTI, the bulk $U(1)^{N_c-1}$ deconfined gauge theory is noninteracting and thus has a unique ground state. For a non-Abelian fTI, the bulk $SU(N_c)$ deconfined gauge theory has a ground state degeneracy\cite{Sato:2007xc} of $N_c$ because $b_1(S^1\times D^2)=1$, i.e., the solid torus has a single noncontractible loop. The $\mathbb{Z}_{N_c}$ fTI also has $N_c$ degenerate ground states on the solid torus coming from the bulk.

\section{Gapless surface states}
\label{SECSURFST}

So far we have considered effective gauge theories for systems with periodic boundary conditions in all spatial directions, i.e., the $3$-torus $T^3$, or for systems with boundaries but where the boundary is gapped since it is described by a CS term. In these two cases, the fermionic partons are gapped everywhere including on the boundary, which allows us to integrate them out. The CS terms break $T$ on the boundary and are absent if $T$ is preserved everywhere. In the latter case, we expect that the fTI should support gapless surface states since each color of partons condenses (at the mean-field level) into a topological band insulator state, which does support gapless surface states. The question therefore arises: what is the nature of the gapless surface state of a fTI?

Since we do not at present have a microscopic model of fTI, i.e., a model of interacting electrons, it is at present difficult to answer this question. Since as we have seen, several different effective gauge theories can give rise to the same quantized fractional $\theta$ angle, we expect a variety of gapless surface states with properties highly dependent on the details of the microscopic model.

From the point of view of the effective gauge theories discussed in this work, the gapless surface states consist of a helical liquid of partons interacting with a three-dimensional gauge field. An effective theory for the $(2+1)$-dimensional surface could be obtained by integrating out the bulk gauge fluctuations. A similar calculation was performed recently\cite{witczak-krempa2010} for the surface helical spinon liquid in spin-charge separated topological Mott insulators,\cite{pesin2010} using a perturbative approach. In this case, the bulk consists of a deconfined $U(1)$ gauge field. Therefore, the results of Ref.~\onlinecite{witczak-krempa2010} should apply qualitatively for the models of Sec.~\ref{SecAbelianModels}, i.e., the deconfined $U(1)^{N_c-1}$ models. Indeed, since the $U(1)^{N_c-1}$ gauge theories are deconfined at weak coupling $g\ll 1$ ($g$ is the parton-gauge boson coupling), we expect that perturbation theory in $g$ should be reliable. Furthermore, Ref.~\onlinecite{witczak-krempa2010} shows that due to the three-dimensional nature of the gauge fluctuations, perturbation theory is better controlled than in the two-dimension case. However, since the microscopic degrees of freedom in a fTI are electrons which correspond to gauge-invariant ``baryon'' operators in the parton gauge theory, one should only calculate correlation functions of gauge-invariant operators. Ref.~\onlinecite{witczak-krempa2010} finds that perturbation theory at 1-loop gives only a logarithmic modification of the tree-level result for the $2k_F$ surface spin-spin correlation function, i.e., $\langle S_+(\b{r})S_-(0)\rangle\sim 1/r^2\rightarrow 1/(r^2\ln k_Fr)$. We therefore expect that the $2k_F$ surface current-current correlation function of the Abelian Coulomb fTI, i.e., its surface electromagnetic response, should also only exhibit logarithmic modification compared to the noninteracting helical Fermi liquid. In the fully gapped Higgs models, the parton-gauge boson coupling should be irrelevant because the bulk gauge fluctuations are massive. Therefore, we expect that the electromagnetic response of the gapless surface state should be the same as that of the noninteracting helical Fermi liquid, up to corrections that are irrelevant at low energies. Finally, for the Coulomb non-Abelian models, since most known examples of these (such as $\mathcal{N}=4$ SYM theory) occur at nonzero coupling $g=g_*$, we expect that the gapless surface state will be a strongly correlated version of the helical Fermi liquid and it is difficult to guess what its properties will be. We conjecture that the resulting state is a ``helical non-Fermi liquid'', and holographic realizations of fTI\cite{HoyosBadajoz:2010ac} may be a useful tool to compute its properties.

\section{Summary}
\label{SECCONCLUSION}

In this work, we considered a variety of gauge theories in $3+1$ dimensions and discussed the conditions they must fulfill to be consistent low-energy descriptions of a fTI. A fTI was defined phenomenologically in previous work\cite{Maciejko:2010tx,Swingle:2010rf} as a $T$-invariant state of interacting electrons which exhibits a quantized fractional axion angle $\theta$ in its low-energy electromagnetic response. We considered Abelian $U(1)$ models and non-Abelian models. In both cases, the confined phase is not an option for a fTI because there would be no fractionally charged excitations in the spectrum, and the existence of fractionally charged states is necessary for a fractional $\theta$ angle to be consistent with $T$. This leaves us with two options: a deconfined (or Coulomb) phase and a Higgs phase. The Coulomb phase of Abelian $U(1)$ models is a theory of noninteracting, gapless gauge bosons. The gaplessness of the gauge bosons does not affect the quantization of $\theta$ because they are electrically neutral. We showed the fractional quantization of $\theta$ explicitly using the Adler-Bell-Jackiw chiral anomaly, which did not require the assumption that the gauge bosons should be gapped. Achieving a deconfined phase in non-Abelian models is more difficult, but non-Abelian models with sufficient electrically neutral gapless matter, such as $\mathcal{N}=4$ SYM theory, are known to realize deconfined phases. These are however usually strongly coupled phases. However, the chiral anomaly still holds in the case of non-Abelian gauge groups, and we could again show explicitly the fractional quantization of $\theta$. Higgs models in which the Abelian $U(1)$ or non-Abelian $SU(N)$ groups were broken down to a discrete $\mathbb{Z}_N$ group were shown to lead to fTI as well, with a fully gapped spectrum in this case.

We investigated the ground state degeneracy of these effective gauge theories on spatial $3$-manifolds of nontrivial topology. On the three-torus $T^3$, the Abelian Coulomb models have a unique ground state. Indeed, this corresponds simply to quantizing several independent flavors of Maxwell electrodynamics in a box with periodic boundary conditions in all three directions. The non-Abelian Coulomb models can have a nontrivial ground state degeneracy on $T^3$ due to the fact that the first homology group $H_1(T^3,\mathbb{Z})=\mathbb{Z}\times\mathbb{Z}\times\mathbb{Z}$ is nontrivial, corresponding to the existence of three inequivalent noncontractible loops in $T^3$. Whether there are multiple ground states or not in the deconfined phase of a $SU(N)$ non-Abelian model depends on whether the center $\mathbb{Z}_N$ of $SU(N)$ is spontaneously broken in the ground state or not. This is a question of dynamics which depends on the details of the model. Higgs models with residual $\mathbb{Z}_N$ gauge group can be viewed as a subset of the previous case. Indeed, $SU(N)$ gauge theories with all fields in the adjoint representation develop the center $\mathbb{Z}_N$ as a global symmetry which can be spontaneously broken in the ground state. Our Higgs models consist of adding adjoint Higgs fields to pure $SU(N)$ gauge theory such that $SU(N)$ is spontaneously broken to its center. Wilson loops around noncontractible loops will still transform nontrivially under $\mathbb{Z}_N$, and their acquiring nonzero expectation values means spontaneous breaking of this global $\mathbb{Z}_N$ symmetry and multiple degenerate ground states. We restricted our consideration of $3$-manifolds $M$ with boundary $\partial M$ to the case $M=\Sigma\times I$ with $\Sigma$ a Riemann surface (say, in the $x,y$ directions) and $I$ an interval (say, in the $z$ direction). In this case, a CS term was induced on the boundary $\partial M=\Sigma\cup\Sigma$, and in the ground state the CS gauge fields on both copies of $\Sigma$ were identified. The resulting CS theory had integer level and its contribution to the total ground state degeneracy (bulk and surface) could be computed using standard methods.

Finally, we briefly commented on what one would expect for the electromagnetic response properties of the gapless surface states based on the general characteristics of the effective gauge theories discussed here. We expect that Abelian Coulomb models should give at most a logarithmically modified version of the noninteracting helical Fermi liquid, while the fully gapped Higgs models should only give corrections that are irrelevant at low energies. The deconfined, strongly coupled non-Abelian models should give rise to the most interesting case. We conjecture that the gapless surface states of non-Abelian fTI are ``helical non-Fermi liquid'' states and suggest that holographic methods\cite{HoyosBadajoz:2010ac} should be a promising way to study their properties.

\acknowledgments

We would like to thank M. Barkeshli, B. A. Bernevig, S. B. Chung, D. J. Gross, D. L. Harlow, J. McGreevy, T. Senthil, S. L. Sondhi, Z. Wang, X. G. Wen, E. Witten, S. Yaida, and M. Yamazaki for useful discussions. This work is funded by the NSF under the grant number DMR-0904264 (SCZ), by the U.S. DOE under grant No. DE-FG02-96ER40956 (AK), by the Stanford Graduate Fellowship Program and the Simons Foundation (JM) and by the Alfred P. Sloan Foundation (XLQ). JM acknowledges the hospitality of the Kavli Institute for Theoretical Physics (KITP) where part of this work was completed.

\appendix
\section{Flat connections and winding numbers}
%JM: I added this appendix to clarify issues for the non-Abelian case on a 3-mfld with boundary.

In the ground state, the allowed gauge field configurations in the $(3+1)$-dimensional spacetime $M\times\mathbb{R}$ are flat connections $a_\mu=U\partial_\mu U^{-1}$. Denote by $a_\mu^\pm=U_\pm\partial_\mu U_\pm^{-1}$ the boundary values of the gauge field, i.e., $a_\mu^\pm\equiv a_\mu(z=\pm L/2)$ and $U_\pm\equiv U(z=\pm L/2)$. We first observe that the $(2+1)$-dimensional boundary gauge fields $a_\mu^+$ and $a_\mu^-$ are related by a gauge transformation, $a_\mu^+=\omega a_\mu^-\omega^{-1}+\omega\partial_\mu\omega^{-1}$ where $\omega=U_+U_-^{-1}$. Therefore, we have
\begin{align}\label{CSgaugetrans}
\frac{1}{4\pi}\int_{\Sigma\times\mathbb{R}}\omega_\mathrm{CS}^+=
\frac{1}{4\pi}\int_{\Sigma\times\mathbb{R}}\omega_\mathrm{CS}^-+2\pi m,
\end{align}
where
\begin{align}\label{CartanMaurer}
m[\omega]=-\frac{1}{24\pi^2}\int_{\Sigma\times\mathbb{R}}\Tr(\omega\mathrm{d}\omega^{-1})^3
\end{align}
is the winding number of $\omega$. In Eq.~(\ref{FFboundaryCS}), the CS level $\Delta\theta_{z=\pm\frac{L}{2}}/2\pi$, which appears in front of the CS terms in Eq.~(\ref{CSgaugetrans}), is half-integer. As a result, $e^{iS}$ in the path integral acquires a factor of $e^{i\pi m[\omega]}$. If gauge transformations $\omega$ with odd winding number $m[\omega]$ were allowed, the path integral would be highly oscillatory and would cancel. In fact, we can show that $m[\omega]$ is necessarily an even integer. Consider the integral of the second Chern form,
\begin{align}\label{integral2ndChern}
\int_{M\times\mathbb{R}}\Tr f\wedge f=\int_{\Sigma\times\mathbb{R}}\omega_\mathrm{CS}^++\int_{\Sigma\times\mathbb{R}}\omega_\mathrm{CS}^-,
\end{align}
since $\partial M=\Sigma\cup\Sigma$. We now substitute $a=U\mathrm{d}U^{-1}$ into Eq.~(\ref{integral2ndChern}). Since this is a flat connection $f=0$, the left-hand side of Eq.~(\ref{integral2ndChern}) vanishes. Since $a^\pm=U_\pm\mathrm{d}U_\pm^{-1}$, we also have $f_\pm=\mathrm{d}a_\pm+ia_\pm\wedge a_\pm=0$ and the CS forms are given by
\begin{align}
a_\pm\wedge\mathrm{d}a_\pm+i\frac{2}{3}a_\pm^3=
a_\pm\wedge(f_\pm-ia_\pm^2)+i\frac{2}{3}a_\pm^3=-\frac{i}{3}a_\pm^3,\nonumber
\end{align}
hence we obtain
\begin{align}\label{zeroChern}
0=-\frac{i}{3}\int_{\Sigma\times\mathbb{R}}\Tr(U_+\mathrm{d}U_+^{-1})^3
-\frac{i}{3}\int_{\Sigma\times\mathbb{R}}\Tr(U_-\mathrm{d}U_-^{-1})^3.
\end{align}
Furthermore, one can show that the winding numbers for a product of gauge transformations add,
\begin{align}\label{windingnumbersadd}
m[\omega_1\omega_2]=m[\omega_1]+m[\omega_2],
\end{align}
as one would expect based on physical intuition. Using Eq.~(\ref{CartanMaurer}) and the cyclic property of the trace, we have
\begin{align}\label{productgaugetrans}
m[\omega_1\omega_2]&=m[\omega_1]+m[\omega_2]\nonumber\\
&\hspace{3mm}+3\int_{\Sigma\times\mathbb{R}}\Tr\left[
(\omega_2\mathrm{d}\omega_2^{-1})^2\wedge\mathrm{d}\omega_1^{-1}\omega_1\right.\nonumber\\
&\left.\hspace{23mm}+
\omega_2\mathrm{d}\omega_2^{-1}\wedge(\mathrm{d}\omega_1^{-1}\omega_1)^2\right].
\end{align}
Using the identity $\mathrm{d}\eta=-\eta\mathrm{d}\eta^{-1}\eta$ for any invertible matrix-valued function $\eta$, we can show that
\begin{align}
\mathrm{d}\left(\omega_2\mathrm{d}\omega_2^{-1}\wedge\mathrm{d}\omega_1^{-1}\omega_1\right)
=&-(\omega_2\mathrm{d}\omega_2^{-1})^2\wedge\mathrm{d}\omega_1^{-1}\omega_1\nonumber\\
&-\omega_2\mathrm{d}\omega_2^{-1}\wedge(\mathrm{d}\omega_1^{-1}\omega_1)^2,\nonumber
\end{align}
hence the last term of Eq.~(\ref{productgaugetrans}) is a total derivative and Eq.~(\ref{windingnumbersadd}) is obtained. Also, using the same identity one can easily show that $m[\eta^{-1}]=-m[\eta]$. For $\omega=U_+U_-^{-1}$ we therefore obtain
\begin{align}
m[\omega]=m[U_+]+m[U_-^{-1}]=m[U_+]-m[U_-],\nonumber
\end{align}
and from Eq.~(\ref{zeroChern}) we have $m[U_+]+m[U_-]=0$, hence
\begin{align}\label{evenwindingnumber}
m[\omega]=2m[U_+]=-2m[U_-]\in 2\mathbb{Z}.
\end{align}

\section{Ground state degeneracy for an Abelian 3D fTI on a $3$-manifold with boundary}

We consider the Abelian $U(1)\times U(1)$ fTI on the spatial 3-manifold $M=T^2\times I$ with $I=[-L/2,L/2]$ in the $z$ direction. Because the bulk 3D emergent gauge fields are free, we can integrate them out explicitly to obtain an effective action for the gauge fields on the 2D boundary $\partial M=T^2\cup T^2$. It is a Maxwell-CS theory with two coupled gauge fields $a^+_\mu$ and $a^-_\mu$ corresponding to the two copies of $T^2$. In the long-wavelength limit $q\ll 1/L$ the two gauge fields become identified $a^+_\mu=a^-_\mu\equiv\alpha_\mu$ and the level of the CS term for $\alpha_\mu$ is the sum of that for the two surfaces, i.e., it is integer.

The spatial manifold is $M=T^2\times I$ with $I=[-L/2,L/2]$. We can always choose the generators of $U(1)\times U(1)$ to satisfy $\Tr t_at_b=\delta_{ab}$, $a,b=1,2$. The action in imaginary time is
\[
S_\mathrm{3D}[a_\mu]=\int d^4x\left(\frac{1}{4g^2}f^a_{\mu\nu}f^a_{\mu\nu}-\frac{i\theta(z)}{32\pi^2}\epsilon_{\mu\nu\lambda\rho}f^a_{\mu\nu}f^a_{\lambda\rho}\right),
\]
with $f_{\mu\nu}^a=\partial_\mu a_\nu^a-\partial_\nu a_\mu^a$, $a=1,2$ the $U(1)\times U(1)$ emergent field strength and
\[
\partial_z\theta=\sum_{\eta=\pm 1}(2k_\eta+1)\pi\delta(z-\eta L/2),
\]
with $k_\eta\in\mathbb{Z}$. To derive an effective 2D action on $\partial M=T^2\cup T^2$, we introduce a Lagrange multiplier which constrains the gauge field to live on $\partial M$. Then we integrate out the bulk gauge field $a_\mu^a$. In other words, we introduce a resolution of unity
\begin{align}
1=\int\mathcal{D}\tilde{a}_\mu^+\mathcal{D}\tilde{a}_\mu^-&
\delta[a_\mu^a(\tilde{x},L/2)-\tilde{a}_\mu^{+,a}(\tilde{x})]\nonumber\\
&\times\delta[a_\mu^a(\tilde{x},-L/2)-\tilde{a}_\mu^{-,a}(\tilde{x})],\nonumber
\end{align}
in the partition function, where $\tilde{a}_\mu^{\eta,a}$, $\eta=\pm$ are two auxiliary gauge fields which are defined only on the 2D surface $\tilde{x}=(x_0=t,x_1,x_2)$, with $\tilde{a}_\mu^{+,a}$ living on the 2-torus at $z=L/2$ and $\tilde{a}_\mu^{-,a}$ living on the 2-torus at $z=-L/2$. We represent the functional delta function as
\begin{align}
&\prod_\eta\delta[a_\mu^a(\tilde{x},\eta L/2)-\tilde{a}_\mu^{\eta,a}(\tilde{x})]=\int\mathcal{D}\tilde{j}^+_\mu\mathcal{D}\tilde{j}^-_\mu\nonumber\\
&\hspace{10mm}\times\exp\left(i\int d^3\tilde{x}\,\tilde{j}^{\eta,a}_\mu(\tilde{x})[a_\mu^a(\tilde{x},\eta L/2)-\tilde{a}_\mu^{\eta,a}(\tilde{x})]\right),\nonumber
\end{align}
where $\tilde{j}^{\eta,a}_\mu$ is a Lagrange multiplier which implements the constraint that $\tilde{a}_\mu^{\eta,a}(\tilde{x})=a_\mu^a(\tilde{x},\eta L/2)$. The idea is to integrate out first $a_\mu$, and then the Lagrange multiplier $\tilde{j}^\eta_\mu$, to get an effective action $S^\mathrm{eff}_\mathrm{2D}$ in terms of the 2D gauge fields $\tilde{a}_\mu^\eta$ alone. In other words, the partition function is
\begin{align}
Z=&\int\mathcal{D}\tilde{a}_\mu^+\mathcal{D}\tilde{a}_\mu^-\mathcal{D}\tilde{j}^+_\mu\mathcal{D}\tilde{j}^-_\mu\mathcal{D}a_\mu
\exp\biggl(-S_\mathrm{3D}[a_\mu]\nonumber\\
&\hspace{10mm}+i\int d^3\tilde{x}\,\tilde{j}^{\eta,a}_\mu(\tilde{x})[a_\mu^a(\tilde{x},\eta L/2)-\tilde{a}_\mu^{\eta,a}(\tilde{x})]\biggr)\nonumber\\
\equiv&\int\mathcal{D}\tilde{a}_\mu^+\mathcal{D}\tilde{a}_\mu^-\,e^{-S^\mathrm{eff}_\mathrm{2D}[\tilde{a}_\mu^+,\tilde{a}_\mu^-]}.
\nonumber
\end{align}
We want to see whether these gauge fields will be identified or not, i.e., whether we obtain two CS theories or only one.

First of all, since the $\theta$-term is a total derivative it contributes only to the boundary piece. Therefore if we write $S^\mathrm{eff}_\mathrm{2D}=S^\mathrm{Max}_\mathrm{2D}+S^\theta_\mathrm{2D}$ we immediately know that
\[
S^\theta_\mathrm{2D}=-i\sum_{\eta}\frac{k_\eta+\frac{1}{2}}{4\pi}\int d^3\tilde{x}\,
\epsilon_{\mu\nu\lambda}\tilde{a}^{\eta,a}_\mu\partial_\nu\tilde{a}^{\eta,a}_\lambda,
\]
i.e., we obtain two decoupled CS terms of half-odd-integer level $k_\eta+\frac{1}{2}$. To compute $S^\mathrm{Max}$ we need to integrate out the 3D bulk gauge fluctuations, which we can do exactly in the Abelian case because the gauge bosons are noninteracting. We can gauge-fix the Maxwell term in the usual way\cite{NagaosaQFT} by adding a $\frac{1}{2g^2\xi}(\partial_\mu a^a_\mu)^2$ term to the Lagrangian. In the Feynman gauge $\xi=1$ the 3D gauge boson propagator is
\begin{align}\label{Deff2D}
D^{ab}_{\mu\nu}(\tilde{k},k_z)=\frac{g^2\delta^{ab}\delta_{\mu\nu}}{\tilde{k}^2+k_z^2},
\end{align}
with $\tilde{k}=(k_0=\omega,k_1,k_2)$. Because the Lagrange multipliers $\tilde{j}_\mu^{\eta,a}(\tilde{k})=\int d^3\tilde{x}\,e^{-i\tilde{k}_\nu\tilde{x}_\nu}\tilde{j}_\mu^{\eta,a}(\tilde{x})$ are independent of $k_z$, we have to sum over all $k_z$ to obtain an effective 2D propagator. The effective 2D Maxwell propagator for the 2D gauge fields $\tilde{a}_\mu^{\eta,a}$ is therefore
\[
D^{ab,\eta\eta'}_{\mu\nu}(\tilde{k})=g^2\delta^{ab}\delta_{\mu\nu}
\frac{1}{L}\sum_{k_z}\frac{e^{-ik_z(\eta-\eta')L/2}}{\tilde{k}^2+k_z^2},
\]
with $|\tilde{k}|=\sqrt{\omega^2+\tilde{\b{k}}^2}$. Because the fields $a_\mu^{\eta,a}$ must satisfy some sort of boundary condition at $z=\pm L/2$ (either Dirichlet or Neumann), $k_z$ is a discrete variable, $k_z=n\pi/L$, $n\in\mathbb{Z}$. Performing the discrete sum over $k_z$, we obtain
\[
D^{ab}_{\mu\nu}(q)=\frac{g^2\delta^{ab}\delta_{\mu\nu}}{|q|\sinh|q|L}
\left(
\begin{array}{cc}
\cosh|q|L & 1 \\
1 & \cosh|q|L
\end{array}\right),
\]
where we denote $q\equiv\tilde{k}$ for simplicity. The inverse propagator is
\begin{align}
[D^{-1}]^{ab}_{\mu\nu}(q)=&\frac{1}{g^2L}\delta^{ab}\delta_{\mu\nu}\frac{|q|L}{|q|\sinh|q|L}\nonumber\\
&\hspace{15mm}\times
\left(
\begin{array}{cc}
\cosh|q|L & -1 \\
-1 & \cosh|q|L
\end{array}\right).\nonumber
\end{align}
To obtain the final form of $S^\mathrm{eff}_\mathrm{2D}$, we need to integrate out the Lagrange multipliers $\tilde{j}_\mu^{\eta,a}$ which simply amounts to inverting the $2\times 2$ matrix propagator (\ref{Deff2D}). Since now all quantities are 2D, we can drop all the tildes for simplicity, and obtain
\begin{widetext}
\begin{align}\label{Seff2D}
S^\mathrm{eff}_\mathrm{2D}[a_\mu^+,a_\mu^-]=&\frac{1}{2g^2L_z}\int\frac{d^3q}{(2\pi)^3}
\frac{|q|L_z}{\sinh|q|L_z}\delta_{\mu\nu}
\left(
\begin{array}{cc}
a^{+,a}_\mu & a^{-,a}_\mu
\end{array}
\right)_{-q}
\left(
\begin{array}{cc}
\cosh|q|L_z & -1 \\
-1 & \cosh|q|L_z
\end{array}
\right)
\left(
\begin{array}{c}
a^{+,a}_\nu \\ a^{-,a}_\nu
\end{array}
\right)_{q}\nonumber\\
&-i\sum_{\eta}\frac{k_\eta+\frac{1}{2}}{4\pi}\int d^3x\,
\epsilon_{\mu\nu\lambda}a^{\eta,a}_\mu\partial_\nu a^{\eta,a}_\lambda,
\end{align}
with $q\equiv(\omega,\b{q})$ and $|q|=\sqrt{\omega^2+\b{q}^2}$. We denoted $L_z\equiv L$ for clarity.

Consider fixing the scaling dimension of the gauge fields by the CS term. The latter is therefore marginal and contains one power of $q$. The effective Maxwell term in Eq.~(\ref{Seff2D}) contains higher powers of $|q|$. Let us expand it to quadratic order in $|q|L_z$,
\begin{align}
S^\mathrm{eff}_\mathrm{2D}[a_\mu^+,a_\mu^-]=&\frac{1}{2g^2L_z}
\int\frac{d^3q}{(2\pi)^3}\delta_{\mu\nu}
\left(
\begin{array}{cc}
a^{+,a}_\mu & a^{-,a}_\mu
\end{array}
\right)_{-q}
\left(
\begin{array}{cc}
1+\frac{1}{3}q^2L_z^2 & -1+\frac{1}{6}q^2L_z^2 \\
-1+\frac{1}{6}q^2L_z^2 & 1+\frac{1}{3}q^2L_z^2
\end{array}
\right)
\left(
\begin{array}{c}
a^{+,a}_\nu \\ a^{-,a}_\nu
\end{array}
\right)_{q}\nonumber\\
&-i\sum_{\eta}\frac{k_\eta+\frac{1}{2}}{4\pi}\int d^3x\,
\epsilon_{\mu\nu\lambda}a^{\eta,a}_\mu\partial_\nu a^{\eta,a}_\lambda.
\nonumber
\end{align}
\end{widetext}
We now consider the long-wavelength, low-energy limit $|q|\ll 1/L_z$. In this limit, the quadratic Maxwell terms $q^2L_z^2$ are irrelevant and the leading term is
\begin{align}
S^\mathrm{eff}_\mathrm{2D}[a_\mu^+,a_\mu^-]=&\frac{1}{2g^2L_z}\int\frac{d^3q}{(2\pi)^3}\delta_{\mu\nu}
\left(
\begin{array}{cc}
a^{+,a}_\mu & a^{-,a}_\mu
\end{array}
\right)_{-q}\nonumber\\
&\times
\left(
\begin{array}{cc}
1 & -1 \\
-1 & 1
\end{array}
\right)
\left(
\begin{array}{c}
a^{+,a}_\nu \\ a^{-,a}_\nu
\end{array}
\right)_{q}\nonumber\\
=&\frac{1}{2g^2L_z}\int d^3x\,(a^{+,a}_\mu-a^{-,a}_\mu)^2.\nonumber
\end{align}
This term contains no derivatives of $a_\mu^{\eta,a}$ but simply implements a constraint. The equations of motion read
\begin{align}
0&=\frac{\delta S^\mathrm{eff}_\mathrm{2D}}{\delta a_\mu^{a,+}}=\frac{\partial L}{\partial a_\mu^{a,+}}=2(a^{a,+}_\mu-a^{a,-}_\mu),\nonumber\\
0&=\frac{\delta S^\mathrm{eff}_\mathrm{2D}}{\delta a_\mu^{a,-}}=\frac{\partial L}{\partial a_\mu^{a,-}}=-2(a^{a,+}_\mu-a^{a,-}_\mu),\nonumber
\end{align}
which imply that $a^{a,+}_\mu=a^{a,-}_\mu$ and the gauge fields on the two 2-tori are identified. Therefore, in the limit $|q|\ll 1/L_z$ the CS term in Eq.~(\ref{Seff2D}) becomes
\begin{align}\label{AppLevelkCS}
S^\theta_\mathrm{2D}=&-i\sum_{\eta}\frac{k_\eta+\frac{1}{2}}{4\pi}\int d^3x\,
\epsilon_{\mu\nu\lambda}a^{\eta,a}_\mu\partial_\nu a^{\eta,a}_\lambda\nonumber\\
=&-i\frac{k}{4\pi}\int d^3x\,
\epsilon_{\mu\nu\lambda}\alpha^a_\mu\partial_\nu \alpha^a_\lambda,
\end{align}
with $\alpha_\mu^a\equiv a^{a,+}_\mu=a^{a,-}_\mu$ and $k=\sum_{\eta}(k_\eta+\frac{1}{2})=k_++k_-+1$ is the effective CS level, which is integer. Therefore, the ground state degeneracy of the $U(1)\times U(1)$ Abelian fTI on $T^2\times I$ is the same as that of the level-$k$ $U(1)\times U(1)$ CS theory (\ref{AppLevelkCS}) on $T^2$, which is known\cite{topdegeneracy} to be $\frac{1}{2}(k+1)(k+2)$.

\bibliography{longfti}

\begin{thebibliography}{150}
\bibitem[{Spi()}]{SpinChargeMott}
\bibinfo{note}{By itself, spin-charge separation in spatial dimensions higher
  than one as it occurs in ``ordinary'' Mott insulators of the spin liquid type
  already signals the existence of a type of topological structure known as
  topological order (sometimes ``Wen-type'' topological order), which is
  characterized by fractionalization and emergent gauge
  structures.\cite{WenBook} In the present context, the word ``topological'' in
  ``topological Mott insulator'' refers to the topology of the single-particle
  bandstructure, sometimes known as ``Thouless-type'' topological order and
  characterized by momentum-space topological invariants and protected boundary
  modes.\cite{ThoulessBook}}.
\bibitem[{Not()}]{NoteSpinMfld}
\bibinfo{note}{The same value applies for all manifolds which are spin, such as
  $T^4$. For a manifold that is not spin the minimal value of the integral of
  $F \wedge F$ is half as large.\cite{Witten1995} As our parton model contains
  fermions and hence can only be formulated on manifolds that are spin, we will
  restrict ourselves to that case.}
\bibitem[{not()}]{noteweights}
\bibinfo{note}{Were it not for the third $U(1)_\mathrm{em}$ entry, the weight
  lattice would be self-dual, i.e. $\mathbf{e}_a=\frac{\mathbf{m}_a}{2\pi}$.
  This would be the case for $U(3)$, or more generally for
  $U(N_c)$.\cite{langlands} We are concerned with the Lie algebra of
  $SU(N_c)\times U(1)_\mathrm{em}$, which is different from that of $U(N_c)$
  because we have chosen a different normalization for the $SU(N_c)$ generators
  than for the $U(1)_\mathrm{em}$ generator. This difference is important, and
  is the mathematical reason for the appearance of $\theta/N_c$ rather than
  $\theta/N_c^2$ in the $\theta$-term for the $U(1)_\mathrm{em}$ gauge field.}
\bibitem[{hol()}]{holonomyvsWL}
\bibinfo{note}{Note that the holonomy $\Gamma(C_i)$ is an element of the gauge
  group $G$, i.e. it is a matrix, whereas the Wilson loop $W(C_i)$ defined in
  Eq.~(\ref{GlobalWL}) is the trace of the holonomy, i.e. it is a $c$-number.}
\end{thebibliography}

\end{document}